\documentclass[twocolumn]{aastex62}
\usepackage{newtxtext,newtxmath}
\usepackage{natbib, amsmath, color}
\usepackage{graphicx, hyperref,lastpage}

\usepackage[T1]{fontenc}

\usepackage{graphicx}	
\usepackage{amsmath}	
\usepackage{amssymb}	
\usepackage{xcolor,ulem}

\shorttitle{Double neutron stars and r-process knees}
\shortauthors{D. Maoz and E. Nakar}

\begin{document}
\title[]{The neutron-star merger delay-time distribution, r-process "knees", and the metal budget of the Galaxy}
\author{Dan Maoz}
\author{Ehud Nakar}
\affiliation{School of Physics \& Astronomy \\ Tel Aviv University, Tel Aviv 69978, Israel}

\begin{abstract}
For a sample of 18 currently known recycled millisecond pulsars (rMSPs) that are in double neutron star (DNS) systems, and 42 rMSPs
with similar properties that are not in DNS pairs, we analyze the  distributions of the characteristic age, $\tau_c$, and the time until merger of the double systems, $\tau_{\rm gw}$. Based on the $\tau_c$ distribution of non-DNS rMSPs, we argue that $\tau_c$ is a reasonable estimator of true pulsar age and that rMSPs are active as pulsars for a long ($\gtrsim$Hubble) time. Among the DNSs there is an excess of young systems (small $\tau_c$) with short life expectancy (small  $\tau_{\rm gw}$)
compared to model expectations for the distributions of $\tau_c$ and $\tau_{\rm gw}$ if, at birth, DNSs have a delay-time distribution (DTD) of the form $\sim\tau_{\rm gw}^{-1}$ (expected generically for close binaries), or for that matter, from expectations from any single power-law DTD. A two-population DNS model solves the problem: the data are best fit by the combination of a "fast" population with DTD going as $\tau_{\rm gw}^{-1.9\pm0.4}$, and a "slow" population of DNSs, with DTD proportional to $\tau_{\rm gw}^{-1.1\pm0.15}$. The fast population can be equivalently represented by a shallow power-law DTD with an exponential cutoff beyond $\tau_{\rm gw}\sim 300$~Myr. The fast population completely dominates, by a factor $A\approx 10-100$, the numbers of DNSs that merge within a Hubble time, and that presumably lead to short gamma-ray bursts (sGRBs) and kilonova explosions. 
Using a simple, empirically based, chemical-evolution calculation, we show that the fast/steep kilonova DTD, convolved with the measured star-formation history of the Milky Way's thick-disk population, naturally reproduces the "knee" structure seen in abundance-ratio diagrams of thick-disk stars, for europium and for two other r-process elements.  As a corollary we show, based again solely on empirical input concerning iron production by supernovae, that the Milky Way is nearly a "closed box" that has retained at least $\sim70-90\%$ of the metals produced over the Galaxy's lifetime. 
\end{abstract}

\vskip 0.5cm



\section{Introduction}\label{sec:intro}
The discovery of the double neutron star (DNS) merger event GW170807 via gravitational waves, and the detection of the ensuing kilonova explosion, were a watershed event for understanding explosive  element production. Analysis of the data lends strong support to the notion that neutron-star mergers lead to a short gamma-ray-burst (GRB), followed by a kilonova that synthesizes of order $0.05$~M$_\odot$ of r-process elements (see \citealt{Nakar2020,Radice2020,Margutti2021} for several reviews). Numerous studies have, since then, addressed the naturally arising followup questions: what is the progenitor population of the merging DNSs? Can the numbers and merger rates of DNSs explain the observed patterns and evolution of r-process abundances in the atmospheres of stars? Do other significant sources of r-process production exist, apart from DNS mergers similar to GW170817 \citep[e.g.,
][]{Cote2017,Hotokezaka2018,Siegel2019,Radice2020,Cowan2021,Siegel2022,Lian2023,Chen2024}

Neutron stars (NSs) are directly observable almost exclusively when they are detected as pulsars, when the Earth is within the rotating pulsar beam during the time that a NS is in its active pulsar phase. Uncertainties concerning pulsar, and pulsar-population, physical parameters, lifetimes, and observational selection effects, have challenged attempts to determine the DNS merger rate based on the known Galactic DNS systems \citep[e.g.,][]{Phinney1991,Curran1995,Kalogera2001,Kalogera2004}, and to compare it to the observed rate of events similar to GW170817 plus its kilonova (further challenged by the fact that this has been, to date, the only case of such a combined detection of gravitational waves plus an electromagnetic counterpart; \citealt{LIGO_BNSRATE23}). As a result, both the predicted and observed rates have been estimated merely to within orders of magnitude (see \citealt{ColomIBernadich2023} for a recent estimate). 

In terms of the potential progenitor populations, DNS systems have been discovered in several types of configurations: in so-called "young pulsars" that are spinning down since their initial formation as high-magnetic-field NS remnants from the core-collapse supernovae of massive stars; in globular clusters (where possibly the two NSs gravitationally captured each other in the dense cluster environment); and in "recycled" or "partially recycled" millisecond pulsars (rMSPs). rMSPs are pulsars that reside in the low period, $P$, and low period derivative, $\dot P$, parameter-space region of the known pulsar population's $P\dot P$ diagram \citep[see][for a review]{Lorimer2008}.
All rMSPs are in binary systems, and a majority of the two-dozen or so currently known DNSs contain a rMSPs. It is generally believed that rMSPs were once neutron stars that spun down (possibly even crossing the pulsar "death line" and ceasing to emit as pulsars), but were then spun up via mass transfer from an evolving companion star. The rMSPs then resumed emitting as pulsars, but with much weaker magnetic fields than young pulsars,
resulting in low luminosities, low $\dot P$, and long lifetimes observable as pulsars. 

Given these characteristics, Galactic DNSs with rMSPs could potentially constitute tracers of the merging DNS progenitor population. A useful 
characteristic of any such population is its delay-time distribution (DTD).
The DTD is the hypothetical distribution of times between the formation, in a brief burst, of a representative stellar population, and the mergers of the DNS systems that have formed from it. A $\sim t^{-1}$ dependence of the DTD is expected for any compact binary merger process that is driven by gravitational-wave losses (e.g. \citealt{Totani2008}), be it DNSs or double white dwarfs (see further below). \cite{BeniaminiPiran2019} have attempted to constrain the DTD of  DNS mergers by analyzing the characteristics of the DNS population, as known at the time of their study, that are not associated with with globular clusters. Their analysed sample included 13 rMSP in DNSs and one DNS with a young pulsar. They assessed that at long delays ($t>1$~Gyr) the DNS-merger DTD follows the expected $t^{-1}$ form, while at earlier times they noted an excess of rapidly merging systems. \cite{BeniaminiPiran2019} estimated that 40\% or more of DNS mergers take place within less than 1 Gyr after system formation. 

The fossil record of metal enrichment visible in the atmospheres of stars may provide further clues to some of the issues above. Among the r-process elements, europium (Eu) abundances have been measured in the largest number of stars. When plotted in traditional chemical-abundance diagrams showing the mass ratio of element X-to-iron, [X/Fe], versus iron-to-hydrogen, [Fe/H] (logarithmic abundance ratios  relative to Solar), stars tend to cluster in a locus often described as a "plateau" of constant [Eu/Fe] from the lowest metallicities, [Fe/H], up to [Fe/H]$\approx -0.5$, where there is a "knee", above which stars tend to have decreasing [Eu/Fe] with increasing [Fe/H]
(e.g. \citealt{Cote2017,Cote2019,Hotokezaka2018}). The [Eu/Fe] locus is reminiscent of 
that for stars in the Galaxy's halo and thick-disk component for the ratios
[$\alpha$/Fe] vs. [Fe/H], where $\alpha$ signifies some of the $\alpha$-elements with mass number that is a multiple of 4, i.e., that can be assembled by adding up consecutive He nuclei to form a nucleus (O, Ne, Mg, Si,...).

The plateau+knee structure in [$\alpha$/Fe] has been traditionally explained in chemical evolution models
as a result of the differing timescales of explosion for different types of supernovae (SNe) that enrich the interstellar gas with different elements, gas which then forms subsequent generations of stars. $\alpha$ elements are produced predominantly in core-collapse SNe, whereas iron comes  from both core-collapse SNe (CC-SNe) and Type Ia SNe (SNe Ia). The massive stars that explode as core-collapse SNe exist only $\lesssim 10$~Myr, much shorter than star-formation and chemical-evolution timescales, and therefore enrichment by CC-SNe is effectively instantaneous, with the CC-SN rate tracking the star-formation rate. SNe Ia, in contrast, are the outcome of thermonuclear combustions of carbon-oxygen white dwarfs (WDs; see \citealt{Maoz2014} for a review of unsolved issues of SN Ia progenitor systems and explosions). In analogy to the DTD of DNSs mentioned above, the DTD of SNe Ia is the distribution of times between the formation of a stellar population and the explosion of some of its members as SNe Ia. Since SNe Ia require the presence of WDs, it is assumed that the value of the DTD of SNe Ia is zero between the brief burst at time $t=0$, and up to an "initial" delay $t_i\sim 40-100$~Myr, simply because this is the minimum time for the formation, through normal stellar evolution, of the first WDs. 
Measurements based on SN Ia rates over the past decade have shown that, at delays $t>t_i$ and up to a Hubble time, the DTD is a monotonically decreasing
function of time close to  $D(t)\sim t^{-1}$ \citep{MaozGraur2017,FreundlichMaoz2021}. Such a DTD is expected (see above) in SN Ia progenitor models where the exploding system consists of two WDs in a close binary, that gradually spiral in and merge through the loss of energy and angular momentum to gravitational waves.  In an environment that experiences  star formation with some rate that depends on time (e.g. in a galaxy), the SN Ia rate as a function of time will be the convolution of the star-formation history with the SN Ia DTD (as defined above, i.e., the "response" of the SN Ia rate to a short star-formation burst). 

Within the above framework, chemical evolution models have often explained the [$\alpha$/Fe] knee as a consequence of the initial delay, $t_i$, until the first SNe Ia appear in a star-forming environment (e.g. \citealt{Weinberg2023}, and references therein). Specifically, for a time $t_i$ after star formation begins, only CC-SNe explode and enrich the gas of subsequent stellar generations, and therefore the [$\alpha$/Fe] ratio, which reflects the weighted mean yields of these elements in the various types of CC-SNe,
remains at a constant "plateau". At times $t>t_i$, the reasoning goes, the SNe Ia "kick in", and their production of iron brings about the monotonic decrease in [$\alpha$/Fe] with time and with increasing overall metallicity [Fe/H]---the "knee". However, in this picture, the very similar knee structure seen for [Eu/Fe]
becomes a puzzle. Gravitational-wave-driven DNS mergers should have a  $\sim t^{-1}$ DTD similar to that of SNe Ia, i.e. they are similarly "delayed" with respect to star formation and to CC-SNe. The Eu from DNS mergers and their kilonovae would then "kick in" at the same time as the Fe from SNe Ia, and no knee in the 
[Eu/Fe] ratio would be expected. 

A number of studies (e.g. \citealt{Hotokezaka2018,Cote2019,Banerjee2020,Tarumi2021}) have confirmed this conflict using detailed chemical evolution models. These studies have shown that a DTD that falls faster with time, e.g., steeper than $t^{-1.5}$, is required in order to produce the observed knee in the [Eu/Fe] ratio. Some authors have proposed that perhaps
DNS-merger kilonovae are not the main or only source of Eu---in addition to DNSs there could be a significant contribution from  some other, "prompt", channel for the production of Eu, such as from special types  (e.g. magneto-rotational) of SNe \citep{Nishimura2017,Mosta2018}, from r-process-element production in outflows during the common envelope phases of NSs and their massive-star companions \citep{GrichenerSoker2019}, from peculiar-magnetar formation \citep{Metzger2008,ThompsonUd-doula2018}, or from collapsars \citep{Siegel2019}. Others have resorted to physical effects to explain the Eu abundance patterns, including: neutron-star natal kicks \citep{Banerjee2020} 
of the DNSs that cause the wider DNS systems (with longer delay) to get 
kicked far from star-forming regions, where they have little
effect on Eu enrichment;
the effects of diffusion through the ISM of the rare r-process elements on the effective delay in their enrichment \citep{Tarumi2021}; or metallicity effects on the DNS merger rates, which in turn are reflected in the DTD \citep{MennekensVanbeveren2014}, effectively steepening it relative to its canonical $t^{-1}$ form.

\citet{MaozGraur2017} measured the SN~Ia DTD by comparing volumetric SN Ia rates as a function of redshift, as measured by field  surveys, to the cosmic
star-formation rate density versus redshift $z$ ("the cosmic star-formation history"; SFH). They then used the DTD, the cosmic SFH, and SN element yield estimates, to plot the mean cosmic [$\alpha$/Fe] vs.[Fe/H] for the Universe as a whole. They noticed  that the cosmic [$\alpha$/Fe] has a "knee" reminiscent of the one seen
in such plots for halo and thick-disk stars in the Milky Way. \citet{MaozGraur2017} realized that the cosmic knee is not the result of any delayed SNe Ia
kicking in. Rather, it is caused by the sharp drop in the cosmic star-formation rate, and the closely tracking CC-SN rate,  following "cosmic noon" at $z\sim2$. 
While the CC-SN rate drops steeply at $z<2$, and with it the production of $\alpha$ elements, the SN Ia rate declines only mildly because it is a convolution of the earlier history with the broad $t^{-1}$ DTD. The cosmic knee is thus an outcome of a decline in $\alpha$-element production, rather than an increase in Fe production.  \citet{MaozGraur2017} argued that the same process, namely a sharp drop in SFH, could be at work behind the [$\alpha$/Fe] knee seen in abundance diagrams for Galactic stars. Using a simple schematic chemical evolution calculation, after some experimentation, they easily found a SFH that reproduces the "high-$\alpha$" sequence of thick-disk and halo stars, with a knee in the observed location in abundance-ratio space. This Galactic SFH consisted of a single short burst of star formation, peaking 11.4 Gyr ago, and lasting several Gyr. 

\cite{Fantin2019} derived the SFH of the thick disk using PANSTARRS1 and {\it Gaia} DR2 data for 25,000 WDs, and found a remarkably similar SFH: a few-Gyr-wide sharp burst about 10~Gyr ago. \cite{XiangRix2022} using LAMOST data for 250,000 subgiant stars, which together with {\it Gaia} eDR3 input provide accurate stellar ages, measured directly the thick disk SFH, finding a single sharp peak in star formation 11.2 Gyr ago, almost exactly the same peak time (although broader in time extent) deduced by \citet{MaozGraur2017} in order to explain the [$\alpha$/Fe] knee . Importantly, the SFH of \cite{XiangRix2022} shows that [Fe/H]$\approx-0.5$ for  the stars that were formed during the peak in the SFH, and therefore strongly suggests that the drop in star-formation rate at later times and higher metallicities
is responsible for the knee in [$\alpha$/Fe] at the same [Fe/H]$\approx-0.5$. Chemical evolution models by \cite{Conroy2022} support this idea. Most recently, \cite{Mason2023} have combined EAGLE hydrodynamical galaxy formation simulations with the VICE
chemical evolution code to conclude that [$\alpha$/Fe] knees occur only in simulated galaxies that encounter a sustained decline in star-formation rate, and only because of the decline, rather than as a result of a delayed 
onset of enrichment by SNe Ia, as previously postulated by \citet{MaozGraur2017}.

In this paper, at first we attempt to constrain the DTD of merging DNS systems based on an observed sample of Galactic DNSs. We then consider whether the DTD that we deduce can reproduce the knee observed in plots of [Eu/Fe] vs. [Fe/H]. The paper is arranged as follows. In \S\ref{sec:pulsars} we compare the rMSPs in Galactic DNS systems to rMSPs that are not in DNS pairs. We find that, for the DNSs, the distributions of system ages and of times-until-merger, when compared to the simplest expectations, are both strongly skewed toward short times, $\lesssim500$~Myr. The observed distributions are well-reproduced by a two-population DNS model, combining DNSs born with a $\sim t^{-1}$ DTD, and a second DNS population born with a $\sim t^{-2}$ DTD (or equivalently, a second population with $t^{-1}$ but with a cutoff at $t \approx 300$~Myr). The combined DTD  that we find predicts that the vast majority ($\sim99\%$) of mergers take place within 1 Gyr. We discuss and compare this result to previous work. In \S\ref{sec:knee} we show how the measured \cite{Fantin2019} SFH for the thick disk, combined with the standard empirically determined DTD for SNe Ia, reproduces the high-$\alpha$ [$\alpha$/Fe] vs.[Fe/H] stellar sequence. We then show that the same SFH, now combined with the DTDs of the two-population model representing DNS mergers (as found in \S\ref{sec:pulsars}), reproduces the 
observed [Eu/Fe] vs.[Fe/H] locus with its knee (and analogous measurements for two additional r-process elements---gandolinium and dysprosium). In \S\ref{sec:discussion} we briefly discuss some physical implications of our results. Finally,
in \S\ref{sec:MW_iron}, we use the empirical inputs of our chemical-evolution calculations to compare the existing iron mass in the Galaxy to the total accumulated iron mass formed by SNe, finding a close match between the two.

\section{Pulsar ages, remaining DNS lifetimes, and the DNS delay-time distribution}\label{sec:pulsars}
With the aim  of identifying the progenitor population of merging DNSs and characterizing their properties, we have defined, from among all the currently known DNSs, a relatively homogeneous sample, namely, all DNSs in which at least one member is a recycled millisecond pulsar. Out of all DNS systems, we exclude the several DNSs associated with globular clusters, and a single DNS (J1906$+$0746) that does not clearly reside in the rMSP region of the $P\dot P$ plane. The 18 sample members are listed in Table~\ref{Table1} (see \citealt{ColomIBernadich2023} for a recent compilation of known DNSs). All of the DNSs in this sample are in the $P\dot P$ region: $17{\rm~ ms}<P<186{\rm ~ms}$; $2 \times 10^{-20}  < \dot{P} < 2 \times 10^{-17}$.

\begin{deluxetable}{lrccrcc}
\tablecaption{Known DNS systems with (partially) recycled millisecond pulsars
\label{Table1}}
\tablehead{
\colhead{ DNS} & \colhead{$P_{\rm orb}$ }&  \colhead{$e$ } &   \colhead{ $\tau_{\rm gw}$} & \colhead{$P_{\rm spin}$} & \colhead{$\dot P_{\rm spin}$ } & \colhead{    $\tau_c$}\\
\colhead{}   &     \colhead{   (days)} & \colhead{}      &  \colhead{          (Gyr)}   &   \colhead{      (ms)}   &     \colhead{ ($10^{-18}$) }   &   \colhead{    (Gyr)}\\}
\startdata
J1946$+$2052    &      0.078  &     0.064  &     0.045    &          17.0    &          0.9   &              0.30 \\         
J1757$-$1854    &      0.184  &     0.606  &     0.076    &          21.5    &          2.6   &              0.13 \\
J0737$-$3039    &      0.102  &     0.088  &     0.086    &          22.7    &          1.8   &              0.20 \\ 
B1913$+$16      &      0.323  &     0.617  &     0.30     &          59.0    &          8.6   &              0.11 \\
J1913$+$1102    &      0.206  &     0.090  &     0.47     &          27.3    &          0.16  &              2.70 \\
J0509$+$3801    &     0.380   &     0.586  &     0.58     &         76.5     &         7.9    &              0.15  \\      
J1756$-$2251    &     0.320   &     0.181  &     1.7      &         28.5     &         1.0    &              0.44 \\
B1534$+$12      &     0.421   &     0.274  &     2.7     &          37.9     &         2.4    &              0.25\\
J1208$-$5936    &      0.632   &    0.348  &     7.2      &         28.7     &        $<0.04$ &             $>11.2$\\
J1829$+$2456    &     1.176    &   0.139   &     55       &          41.0    &          0.053 &              12.3 \\
J1759$+$5036    &      2.043   &   0.308   &    177       &         176.0    &          0.243 &              11.5 \\
J1325$-$6253    &      1.816   &     0.064 &     189      &          29.0    &          0.048 &              11.3 \\
J1411$-$2551    &     2.616    &   0.170   &    466       &            62.5  &          0.096 &              10.3 \\
J0453$+$1559    &     4.072    &   0.113   &   1450       &           45.8   &           0.19 &              3.90 \\
J1811$-$1736    &     18.779   &   0.828   &   1800       &        104.2     &         0.90   &              1.83 \\
J1518$+$4904    &     8.634    &   0.249   &   8840       &         40.9     &         0.027  &              23.8 \\
J1018$-$1523    &   8.984     &  0.228   & $10^4 $&  83.152    &       0.11     &    12.0           \\
J1930$-$1852    &    45.060    &  0.399   &$5\times 10^5$ &       185.5      &        18.0    &              0.16 \\
\enddata
\tablecomments{DNS and rMSP properties: binary orbital period, eccentricity, and time until merger ($P_{\rm orb}$, $e$, and $\tau_{\rm gw}$, respectively); rMSP spin period, period derivative, and characteristic age ($P_{\rm spin}$,  $\dot{P}_{\rm spin}$, and $\tau_c$, respectively.}
\end{deluxetable}

\begin{figure}
    \centering
    \includegraphics[width=9cm]{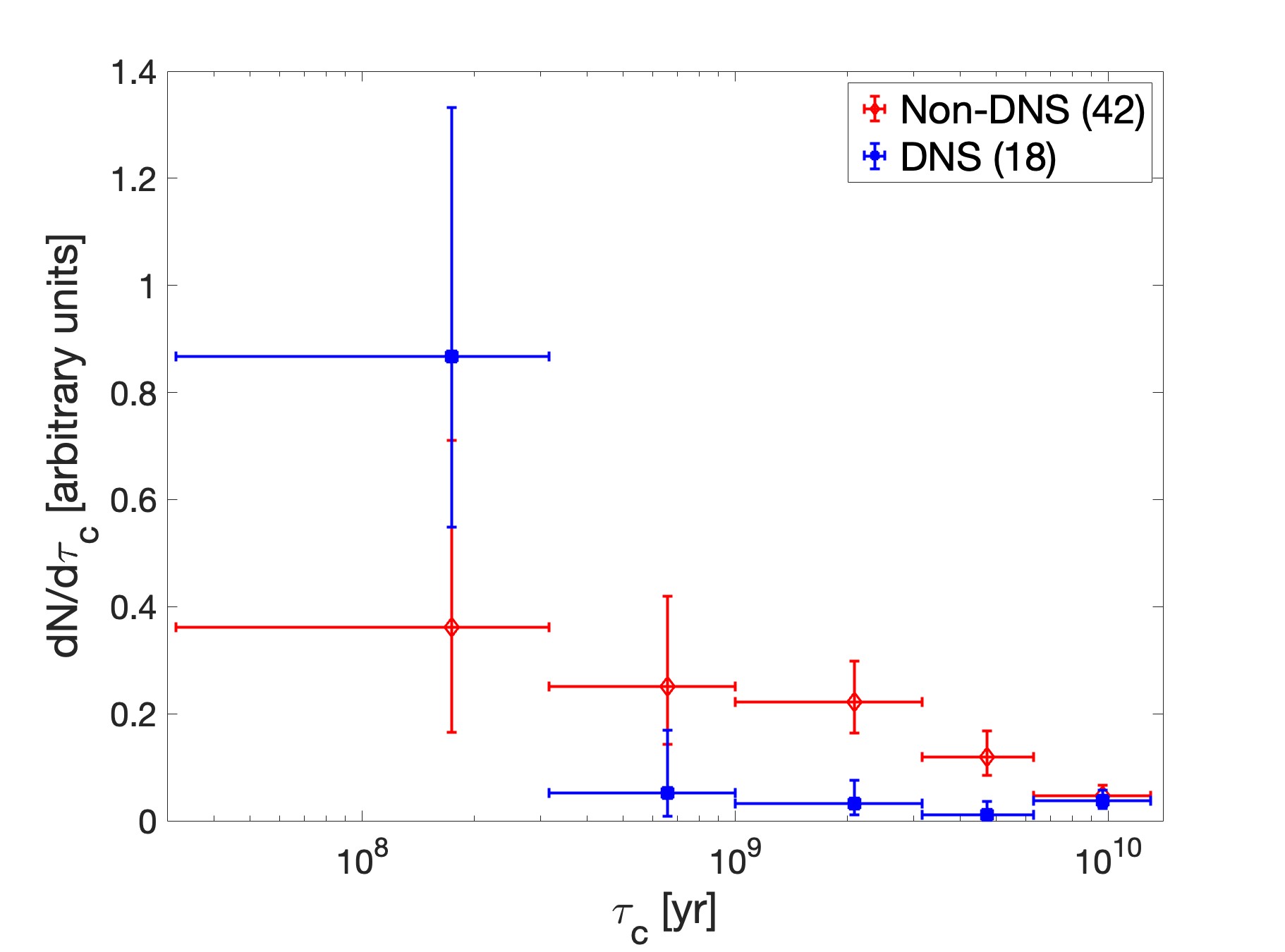}
    \caption{Distribution of characteristic pulsar "ages", $\tau_c\equiv P/(2\dot P)$, per linear interval of $\tau_c$, for two samples of Galactic rMSP systems: 18 rMSP systems that consist of DNSs (blue squares, Table~\ref{Table1}); and (red diamonds) all 42 rMSPs having an identified binary companion to the pulsar that is not a neutron star, and spin period and period derivative in the same region of the $P \dot P$ plane as the 18 DNSs. The figure suggests (see text for details) that: (i) $\tau_c$ is a reasonable estimator of the true pulsar age; (ii) rMSPs live for a duration that is at least comparable to the Hubble time; and (iii) there is a population of young ($<300$ Myr) pulsars that dominates
    among the DNS systems. A small number of systems (one in the DNS sample and two in the non-DNS sample) with $\tau_c$ larger than the age of the universe by a factor of order unity, have been included in the last age bin ($6.3-13$ Gyr). rMSPs in globular clusters have been excluded from both samples. Pulsar data are from the ATNF Pulsar Catalog, {\tt https://www.atnf.csiro.au/research/pulsar/psrcat/} \citep{ATNF2005}.
    }
    \label{fig:dN_dtauc}
\end{figure}

Pulsars (of all types, not just those in DNS systems) are often assigned a "characteristic age", $\tau_c\equiv P /(2\dot P)$, also known as the spin-down age, which is expected to be a rough measure of the time elapsed since the formation of the pulsar. There are known sources of uncertainty of this age estimate. First, the estimate is based on  assuming a braking index of 3 (appropriate for braking by a dipolar  field), whereas the few pulsars with measured indices display a range of values \citep[][and references therein]{Lyne2015}. This will typically introduce an error of order unity, or perhaps up to a factor of a few, to the age. Second, the secular acceleration due to the transverse velocity between the pulsar and the observer leads to an overestimate of $\dot P$, and hence an underestimate of a  pulsar's age \citep{Shklovskii1970}. This effect can be corrected if the distance and transverse velocity of a pulsar are known, and the correction is typically small (less than a factor 2) for pulsars in the $P\dot P$ region of our DNS sample \citep{Kiziltan2010}. For example, for five DNSs in Table~\ref{Table1}, for which distances and transverse velocities have been measured, this correction is $<10\%$. Finally, the spin-down age estimate assumes that $P$ is much larger than the initial pulsar period, but if this assumption is wrong then the pulsar could be much younger than $\tau_c$. This could introduce a significant age correction for very rapidly rotating pulsars, with $P\lesssim 10$ ms, but for pulsars such as those in DNS systems the correction is again expected to be of order unity \citep{Kiziltan2010}. 
In addition to these known sources of uncertainty in the age, there could potentially be changes in magnetic field strength or structure, on time scales shorter than the pulsar age, which would again distort $\tau_c$ as an age estimator, but little is known about this possibility. Altogether,  we conclude that  $\tau_c$ could constitute a reasonable estimator, to better than an order of magnitude, of the age of rMSPs, particularly in the DNS region of the $P\dot P$ plane, but this hypothesis requires additional tests.  

To explore the issue, in Fig.~\ref{fig:dN_dtauc} we plot the distribution of $\tau_c$ for the DNS sample; and for 42 known rMSPs with a companion  that is not a NS (in all cases a WD), and that reside in the same region in the  $P \dot P$ plane as DNSs, namely $10~{\rm ms}<P<200~{\rm ms}$ and $10^{-20}<\dot P<3\times 10^{-17}$. The DNS pulsars and these non-DNS pulsars should plausibly be similar in their physical characteristics.
Several features are apparent. In both samples there is a cut-off in rMSP numbers at $\tau_c$  values near the age of the universe. Seven of the 18 DNSs and 13 of the 42 non-DNS systems have $\tau_c>5$ Gyr, but only one DNS and two non-DNS systems have $\tau_c$ longer than the age of the universe (yet still smaller than twice the Hubble time). This already argues  that $\tau_c$ is a good estimator of the true pulsar age (to within a factor of 2 or so). Furthermore, it suggests that many pulsars are active, and thus detectable, for a duration that is at least as long as the Galaxy's age.  

If indeed rMSPs have long lifetimes as active pulsars, and $\tau_c$ is a reasonable estimator of rMSP age, then, for a roughly constant star-formation history of the Milky Way's thin disk \citep{Cukanovaite2023}, to which population most MSPs belong \citep{Gaspari2024}, one would expect a roughly uniform distribution of rMSP ages, per linear age interval $dN/d\tau_{\rm c}$, up to some cutoff corresponding to the age of the Milky Way.  
Fig.~\ref{fig:dN_dtauc} shows that the $\tau_c$ distribution is indeed extended for both samples. For the non-DNS sample, excluding the highest-age bin (which includes a period of several Gyr before star formation began in the thin disk), the distribution seems to fall  mildly with age, but is consistent with being flat or nearly flat. This provides additional support to the notion of longevity of rMSPs and to the approximate reliability of $\tau_c$ as an age estimator.
The $\tau_c$ distribution  of the DNSs in Fig.~\ref{fig:dN_dtauc}, however, is different. It shows a large excess of systems at short ages, $\tau_c<300$~Myr, suggesting an over-representation  of young systems among rMSPs that are in DNSs. In other words, if $\tau_c$ is a reasonable estimator of pulsar age, then the non-DNS $\tau_c$ distribution suggests that rMSPs have long (of order Hubble-time) lifetimes as active pulsars. The $\tau_c$ distribution of the DNS sample then argues that, as opposed to the non-DNS systems, many DNSs (though not all) have very short lifetimes. We return shortly, below, to the details of the DNS distribution of $\tau_c$.


A complementary timescale that we explore next, and that can be estimated (accurately) for all the DNSs, is the time $\tau_{\rm gw}$ until the merger of each system, driven by the loss of energy and angular momentum to gravitational waves. 
As opposed to the approximate age, $\tau_c$,  i.e. the time since birth of a pulsar, $\tau_{\rm gw}$ is the precise time until death of a DNS system, determined by its constituent masses, separation, and eccentricity. Table~\ref{Table1} lists $\tau_{\rm gw}$, by increasing order,  for the DNS sample. Remarkably, the youngest DNS pulsars (with the lowest $\tau_c)$, which constitute the "youthful excess" in Fig.~\ref{fig:dN_dtauc}, tend to also have  the lowest values of time--till-merger $\tau_{\rm gw}$. 
At first sight, it is odd that the two timescales are correlated when, at face value, they are physically unrelated (the timescale over which one of the pulsars in a DNS system slows down due to internal neutron-star processes, versus the time until two neutron stars merge because of gravitational wave losses.) We will show, below, that the effect is a result of the particular DTD of DNSs.

The suggestion, by  Fig.~\ref{fig:dN_dtauc}, of a steady production rate of rMSPs over the Galaxy's history, along with a long lifetime as an active pulsar for each rMSP, raises the possibility of constraining the delay-time distribution of kilonovae, $D_{\rm kn}(t)$, directly from the observed distribution of $\tau_{\rm gw}$ of the DNS sample. A caveat to this is that the observed distribution may have been 
strongly distorted by selection effects. We examine this question in Appendix A, and conclude that it is unlikely that selection effects play a major role  in the $\tau_{\rm gw}$ distribution, apart from possibly excluding from the sample systems with $\tau_{\rm gw} \lesssim 50$ Myr. 
 Assuming that rMSPs in DNS systems are the progenitor population of kilonovae, different forms of $D_{\rm kn}$ will lead to different distributions of $\tau_{\rm gw}$. To see this, we adapt to the present context the analysis by \cite{MaozBadenesBickerton2012} of the evolution with cosmic time of the separation distribution of a population of binaries that are inspiraling via gravitational wave losses.

Suppose a single population of DNSs forms at time $t=0$ with some distribution of initial separations, which entails a distribution of times-till-merger (or "delay times"), $D'(\tau')$. Systems with delays in the range $\tau'$ to $\tau' +d\tau'$ will migrate, after a time $t$, to a delay bin $\tau$ to $\tau+d\tau$ in the evolved distribution $D(\tau,t)$, where the transformation from $\tau'$ to $\tau$ is simply $\tau=\tau'-t$, resulting from the shortened delay after the elapsed time $t$, and $d\tau=d\tau'$. Conservation of the number of systems requires that 
\begin{equation}
D(\tau, t) d\tau =D'(\tau') d\tau' ,
\end{equation}
and therefore the distribution of delay times at time $t$ is
\begin{equation}
D(\tau,t)=D'(\tau + t),
\end{equation}
i.e. just the original distribution but shifted to earlier delays by $t$.
If we now consider a series of DNS populations, each with the same initial DTD, $D'(\tau')$, tracking the star-formation rate ${\rm SFR}(t)$ between $t=0$ and the present age of the Galactic disk, $t_0$, then the present-day distribution of delay times is 
\begin{equation}\label{eq.taugw}
\frac{dN}{d\tau_{gw}} = \int_{0}^{t_0} {\rm SFR}(t_0 -t) D(\tau_{gw}, t) dt .
\end{equation}

Let us further suppose, for instance, that the initial DTD is a power law of index $\alpha$, $D'(\tau')\propto\tau'^{\alpha}$, and has an "initial delay" $t_i$, analogous to the initial delay already mentioned in the context of SNe Ia and double-WD mergers, in \S~\ref{sec:intro}. In the current context, $t_i$ is the total time elapsed between the formation of a stellar population and the occurrence of the first DNS mergers from that population.\footnote{\label{tifootnote} Strictly speaking, there are several, subtly different, initial delays in the problem: the time $t_{i, {\rm dns}}$, between star-formation and the formation of the first DNS systems that are "ready" to begin their inspiral toward merger (set by the stellar and binary evolution timescales of binary stars that end up as DNS rMSPs); the time $t_{i, {\rm gw}}$ between DNS formation and the first DNS merger events (set by the minimum separation and maximum eccentricity of newly born DNSs, leading to the shortest $\tau_{\rm gw}$); and the sum of these two timescales, $t_i$. In Appendix B, we elaborate on the different initial delays
and on where and how each should, in principle, be used. In the rest of this paper, however, the maximal value that we consider for $t_i$ (the longest of the three initial timescales), $t_i=10$~Myr, is still much smaller than the smallest observed value of 
$\tau_{\rm gw}$, making our results insensitive to the choice of the type of initial delay. We therefore do not distinguish among the three different types of initial delay.}
Then, for $t>t_i$, $ D(\tau, t)\propto (\tau + t)^{\alpha}$. For an assumed constant star-formation rate in the Galactic disk, ${\rm SFR}(t)={\rm const.}$, over the past $t_0=8$~Gyr, we then have for the present-day distribution of times-till-merger 
\begin{equation}
\label{eq.taugwdist1}
    \frac{dN}{d\tau_{\rm gw}} \propto \ln\frac{\tau_{\rm gw}+t_0}{\tau_{\rm gw}}, ~~~~~~~{\rm for} ~\alpha=-1,
\end{equation}
and
\begin{equation}
\label{eq.taugwdist2}
  \frac{dN}{d\tau_{\rm gw}} \propto  \tau_{\rm gw}^{\alpha+1} - (\tau_{\rm gw} + t_0)^{\alpha+1}  ~~~{\rm for}~\alpha \neq -1.
\end{equation}
The life-expectancy distribution  $dN/d\tau_{\rm gw}$ is, under these assumptions, essentially a broken power law with smooth-transition break
at $\tau_{\rm gw}=t_0$, and with indices $\alpha$ at $\tau_{\rm gw}\gg t_0$ and  $\alpha+1$ at $\tau_{\rm gw}\ll t_0$. This recalls the broken power-law distribution of the binary separations, shown by \cite{MaozBadenesBickerton2012}.

\begin{figure}
    \centering
    \includegraphics[width=9cm]{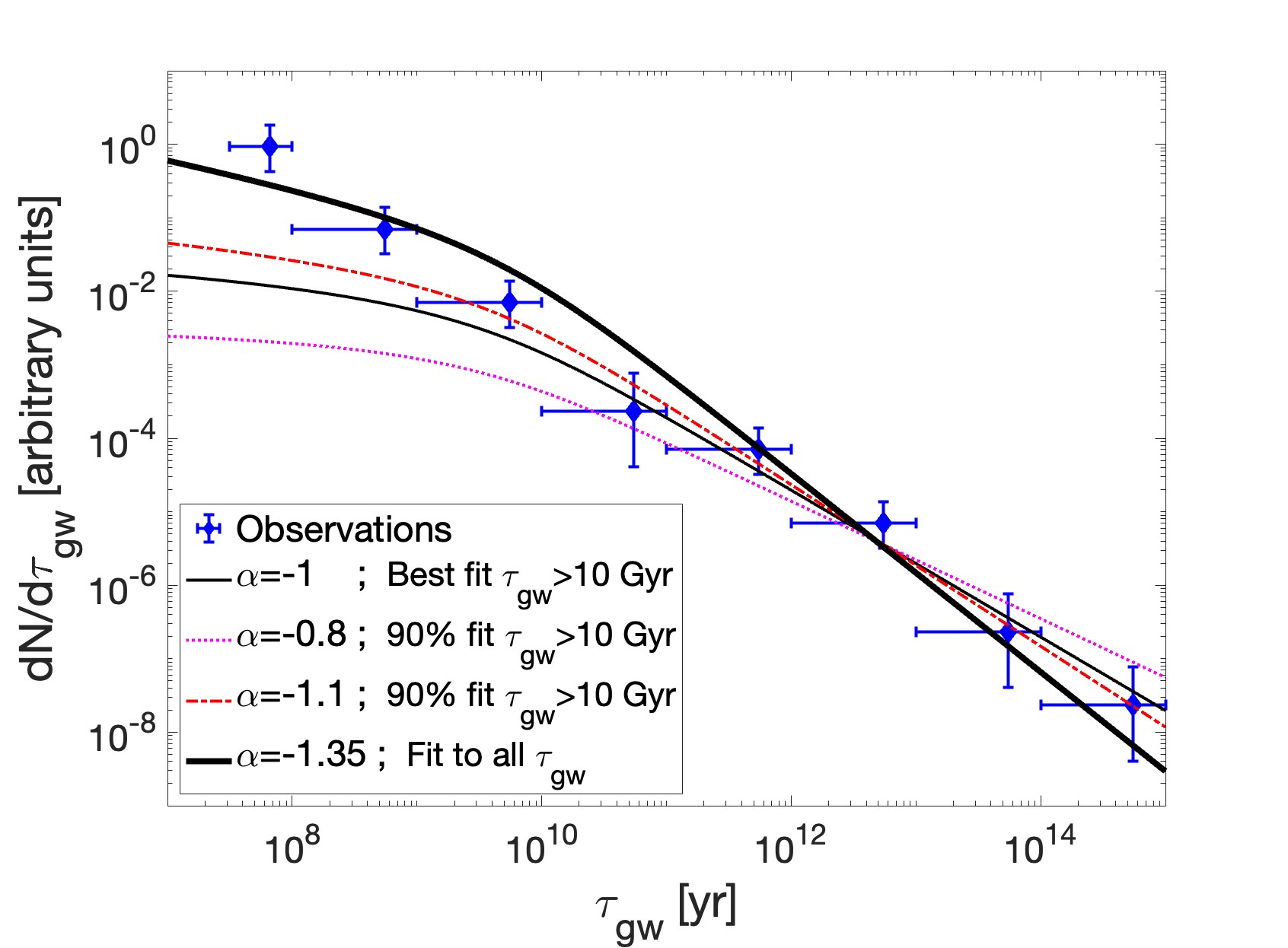}
    \caption{Observed distribution (crosses) of time-till-merger, $\tau_{\rm gw}$ for the 18 DNS systems in Table~\ref{Table1}, and four
    predicted model distributions (curves, as labeled, based on Eqns.\ref{eq.taugwdist1}-\ref{eq.taugwdist2}) for different values of the power-law index $\alpha$ in the delay-time distribution
    $D(\tau_{\rm gw},\alpha)$ of DNS mergers. The three thin curves shown correspond to the maximum likelihood fit ($\alpha=-1.0$) to the observed distribution of $\tau_{\rm gw}$ at  $\tau_{\rm gw}>10$~Gyr, and to the 90\% confidence interval. The thick curve is a model within the $1\sigma$ range of the best fit to the data over all values of $\tau_{\rm gw}$. The broken-power-law form of all the models does not resemble the data. As a result, all such models
    underpredict the number of DNS systems at short $\tau_{\rm gw}$, and the model fit over the full range of  $\tau_{\rm gw}$ also overpredicts the number of systems at intermediate $\tau_{\rm gw}$. This indicates that at least two populations of DNSs, with different delay-time distributions, are required in order to reproduce the observed distribution of  $\tau_{\rm gw}$. The models shown are fit to the actual observed values of $\tau_{\rm gw}$ (see Appendix C for the fitting method), rather than to any cumulative or binned distribution of $\tau_{\rm gw}$, shown here for illustrative purposes only.
    }
    \label{fig:dN_dtaugw}
\end{figure}
Figure~\ref{fig:dN_dtaugw} shows $dN/d\tau_{\rm gw}$ for the DNS sample. Using a maximum-likelihood fit to the data, described in Appendix C,
we have found (curves in Fig.~\ref{fig:dN_dtaugw}) the best-fitting model distributions, as given by Eqns.~\ref{eq.taugwdist1}-\ref{eq.taugwdist2}, for a single power-law DTD, varying the values of the index $\alpha$. We show models both for fits to all $\tau_{\rm gw}$, and only to the data with $\tau_{\rm gw}>10$ Gyr, which are expected to behave as a single power law. 
As noted above, the predicted distribution of $\tau_{\rm gw}$, for DNS populations having a DTD that is a single power law, is always a broken power-law, where the break is around the Galaxy's age. In contrast, 
the observed distribution of $\tau_{\rm gw}$, shown in Fig.~\ref{fig:dN_dtaugw}, is remarkably close to a single power law, which none of the single power-law DTD distributions predicts. As a result, no matter what range of $\tau_{\rm gw}$ in the data is fit, the models cannot satisfactorily reproduce the 
observed distribution. Specifically, for the fits to long $\tau_{\rm gw}$ shown in the figure, there is a large excess, over the model expectations,
of systems with short times until merger. At $\tau_{\rm gw}<100$~Myr, the excess is by at least an order of magnitude. For example, integrating Eq.~\ref{eq.taugwdist2} for the steepest power-law model within the 90\% confidence interval ($\alpha=-1.13$) over $10~{\rm Myr}<\tau_{\rm gw}<100$~Myr (with the model normalized to produce 18 systems over the entire observed range of $\tau_{\rm gw}$), we would expect 0.21 DNSs with $10~{\rm Myr}<\tau_{\rm gw}<100$~Myr, yet we observe three such systems in our sample.
The Poisson probability for this is $P(N \geq 3|\lambda=0.21)=0.001$ and thus this model can formally be rejected at high significance.  
If we fit a single power-law DTD model to the data over the full $\tau_{\rm gw}$ range, then the best fit model (thick curve in Fig.~\ref{fig:dN_dtaugw}) has $\alpha \approx -1.3$. This power-law index value is, in a sense, a compromise between the one required to fit the data at $\tau_{\rm gw} \gg t_0$ and the one required at  $\tau_{\rm gw} \ll t_0$. This model nonetheless still underpredicts the number of DNSs with $10<\tau_{\rm gw}<100$~Myr to be 0.55 (instead of 3 observed).
We conclude that the agreement of a single-power-law DTD model with the data is, at best, marginal, and it is worthwhile to look for models that provide a better match.

\begin{figure}
    \centering
    \includegraphics[width=9cm]{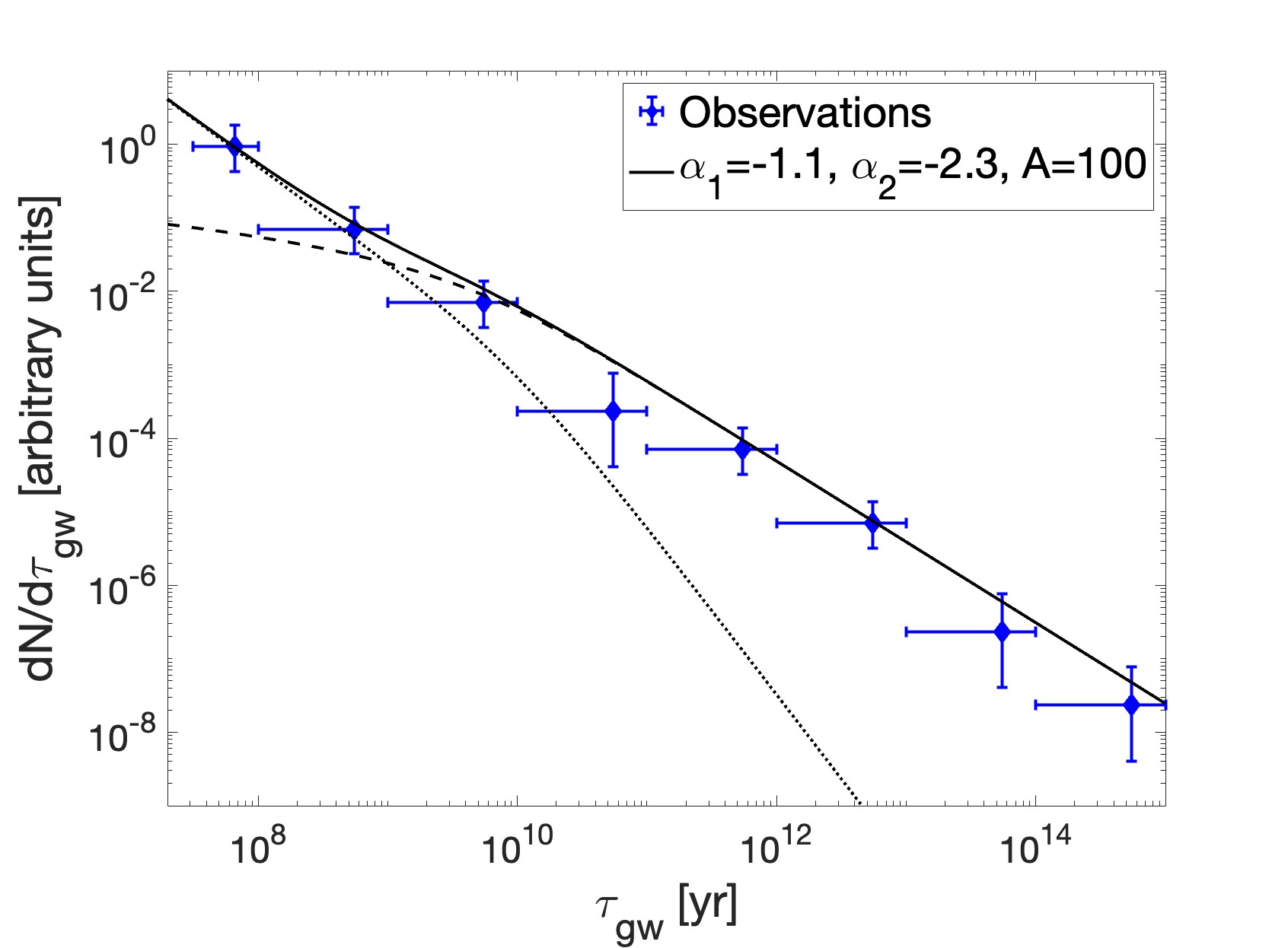}
    \includegraphics[width=9cm]{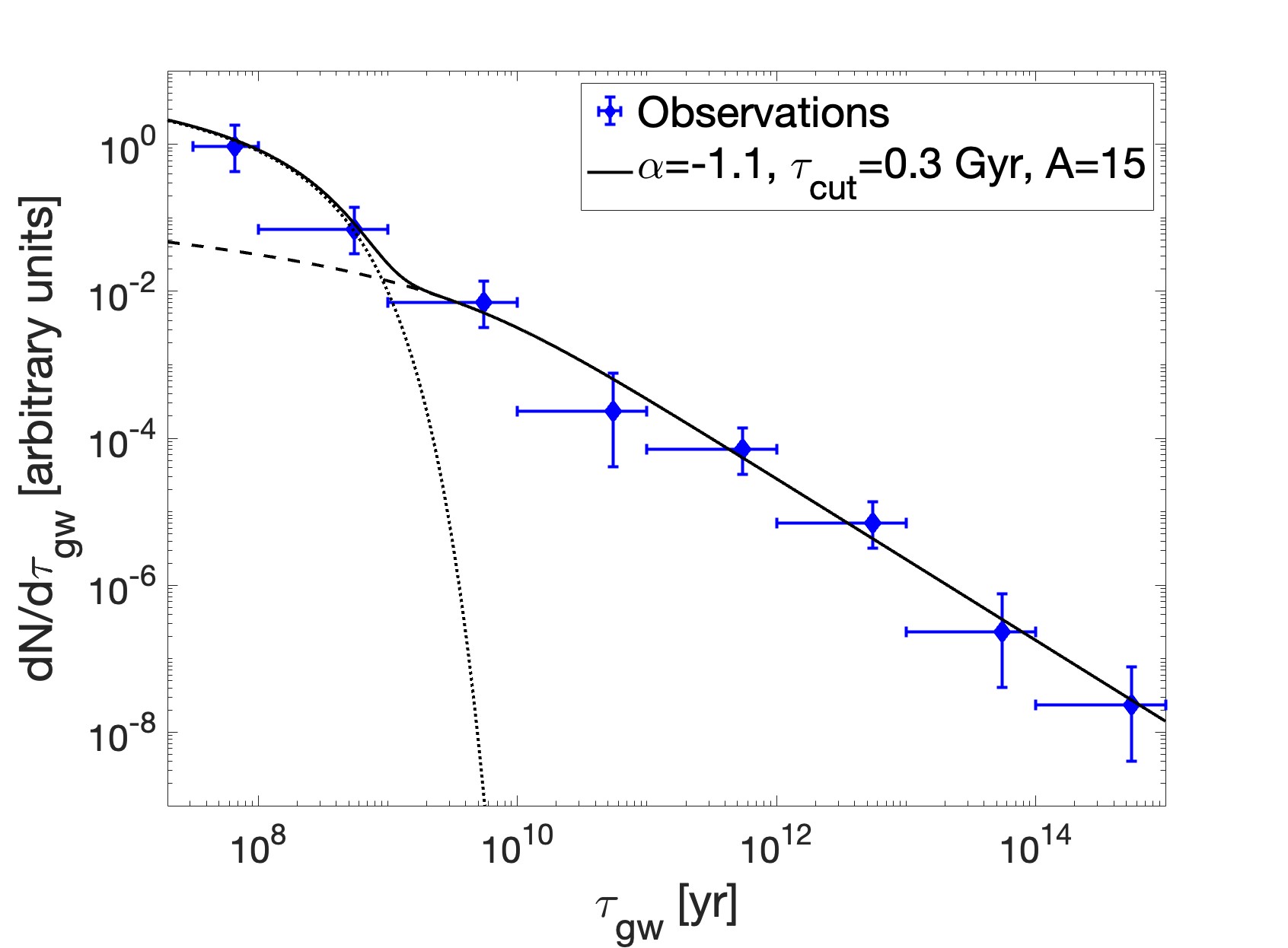}
    \caption{Observed distribution (crosses) of time-till-merger, $\tau_{\rm gw}$ for the DNS sample (same as Fig.~\ref{fig:dN_dtaugw}), and two-population DNS model distributions that fit the data (solid curves). Dashed and dotted curves show the contributions of the individual populations. {\it Upper panel:} Two populations with power-law DTDs with indices as marked, and ratio $A$ of integrated numbers up to a Hubble time. {\it Lower panel:} One population with a DTD that is a power law $\tau^\alpha$, and another with a DTD with an exponential cutoff, $\tau^{-1} \exp(-\tau/\tau_{\rm cut})$. Ratio $A$ is defined the same as above. The two-population models, both of them having a dominant "prompt" DNS-merger DTD component, reproduce well the observed $\tau_{\rm gw}$ distribution. 
    }
    \label{fig:dN_dtaugw2PL}
\end{figure}

To proceed, we have attempted to fit a two-population model, each population with a power-law DTD with a different index,  to the observed $\tau_{\rm gw}$ distribution. Using a Markov-Chain Monte-Carlo (MCMC) procedure, we have tested the likelihood of a family of models with parameters $\alpha_1$, $\alpha_2$, and $A$, consisting of one population of DNSs that are born with a DTD proportional to $\tau^{\alpha_1}$, a second population with DTD proportional to $\tau^{\alpha_2}$, and a  ratio $A$ between the integrals over time of the two DTDs, from $t_i=10$ Myr to a Hubble time. Details of the MCMC calculation and its results, as well as of the (minor) effect of the choice of $t_i$, are in Appendix C.
Fig.~\ref{fig:dN_dtaugw2PL}  (upper panel) shows again
the observed $dN/d\tau_{\rm gw}$ distribution for the 18 DNS systems, now compared to a two-power-law model with parameters close to the most likely ones.  Unsurprisingly, the predicted model distributions from each of the two populations, each of which is a broken power law with a smooth break at $t_0=8$~Gyr,
can easily combine to produce a single power law, as observed for the data (and which, somewhat paradoxically, cannot 
be well reproduced by a single population with a single-power-law DTD, which always predicts a broken-power-law observed $dN/d\tau_{\rm gw}$). The best-fit parameters and $1\sigma$ uncertainties
are $\alpha_1=-1.1\pm 0.15$, $\alpha_2=-1.9 \pm {0.4}$, and $\log_{10}(A)=1.95^{+0.7}_{-0.9}$. We note that the uncertainties quoted here and elsewhere in the paper are statistical and do not include systematics (which most likely dominate the uncertainty), such as  unaccounted-for selection effects that most likely {\it reduce} the observed number of system with small $\tau_{\rm gw}$ (see Appendix A). In this two-population picture, the merger rate is strongly dominated by short-lived systems. For example, in the best-fit model, $\sim 99\%$ of mergers are of systems with time until merger $<1$~Gyr, while only $\sim 1\%$ are of systems with longer life expectancies.

In addition to the two-power-law DTD model above, we have also experimented with a model consisting of two populations with power-law DTDs, but with one of the power laws having an exponential cutoff with timescale $\tau_{\rm cut}$. Explicitly, the first population has DTD~$\propto \tau^{\alpha}$, and the second population has a DTD~$\propto \tau^{-1} \exp(-\tau/\tau_{\rm cut})$. The free parameters of this model are $\alpha$, $\tau_{\rm cut}$ and a relative normalization factor $A$, defined as in the two-power-law model.
As seen in Fig.~\ref{fig:dN_dtaugw2PL} (lower panel) this two-population, exponential-cutoff, model is equally effective in explaining the observed $\tau_{\rm gw}$ distribution. From the MCMC calculation (see Appendix C) the best-fit parameters and $1\sigma$ uncertainties for this model 
are $\alpha=-1.1\pm 0.1$, $\log_{10}(\tau_{\rm cut}/{\rm yr})=9^{+0.4}_{-0.5}$, and $\log_{10}(A)=1.25^{+0.5}_{-0.8}$.
In this model too, the merger rate is dominated by short-lived systems, although somewhat less than in the two-power-law model---in the best fit version, 93\% of the mergers are by systems with $\tau_{\rm gw}<1$~Gyr.
Based on the Akaike Information Criterion, there is a 94\% probability that the more-complex two-population models (each with three free parameters) are preferred over the single power-law model (with its one parameter), and therefore the improved fit of both of the two-population models  justifies the additional complexity introduced. 

Considering the results above, we now return to the observed distribution of $\tau_c$, the characteristic age of the DNS sample,
to investigate whether or not it is consistent with the pictures of one or two DNS populations with the indicated power-law DTDs. Let us assume, again, that rMSPs have long lifetimes ($\gtrsim$~Hubble time), and that $\tau_c$ is a reliable indicator of age. 
From among all the DNSs of age $\tau_c$, i.e., DNS systems that were formed a time $\tau_c$ ago, we can see only those that, at formation time, had $\tau_{\rm gw}>\tau_c$, since those with shorter merger times have merged and disappeared. The predicted distribution of ages $\tau_c$ is  thus
\begin{equation}\label{eq.taucdist}
\frac{dN}{d\tau_c} = {\rm SFR}(t_0 -\tau_c)\int_{\tau_c}^{\tau_{\rm max}}  D(\tau) d\tau ,
\end{equation}
where $\tau_{\rm max}$ is the maximal value of $\tau_{\rm gw}$ possible, either because of a cutoff in the DTD (e.g. if DNSs cannot remain 
bound beyond some separation) or perhaps because it becomes difficult to detect DNSs as such beyond some orbital period. For our purposes, we will assume for $\tau_{\rm max}$
the largest  $\tau_{\rm gw}$ observed in the sample, $5\times 10^5$~Gyr. For a constant star formation history, it is straight-forward to calculate  $dN/d\tau_c$ for the various DTD models discussed above. 
For example, for a single DNS population with DTD proportional to a power law, $\tau^\alpha$, the age distribution is 
\begin{equation}
\label{eq.taucdist1}
    \frac{dN}{d\tau_c} \propto \ln\frac{\tau_{\rm max}}{\tau_c}, ~~~~~~~{\rm for} ~\alpha=-1,
\end{equation}
and
\begin{equation}
\label{eq.taucdist2}
  \frac{dN}{d\tau_c} \propto \tau_c ^{\alpha+1} - \tau_{\rm max}^{\alpha+1}   ~~~{\rm for}~\alpha \neq -1.
\end{equation}
For the power-law DTD with an exponential cutoff, the integral in Eq.~\ref{eq.taucdist} has no analytic solution, but it is easily evaluated numerically.

\begin{figure}
    \centering
    \includegraphics[width=9cm]{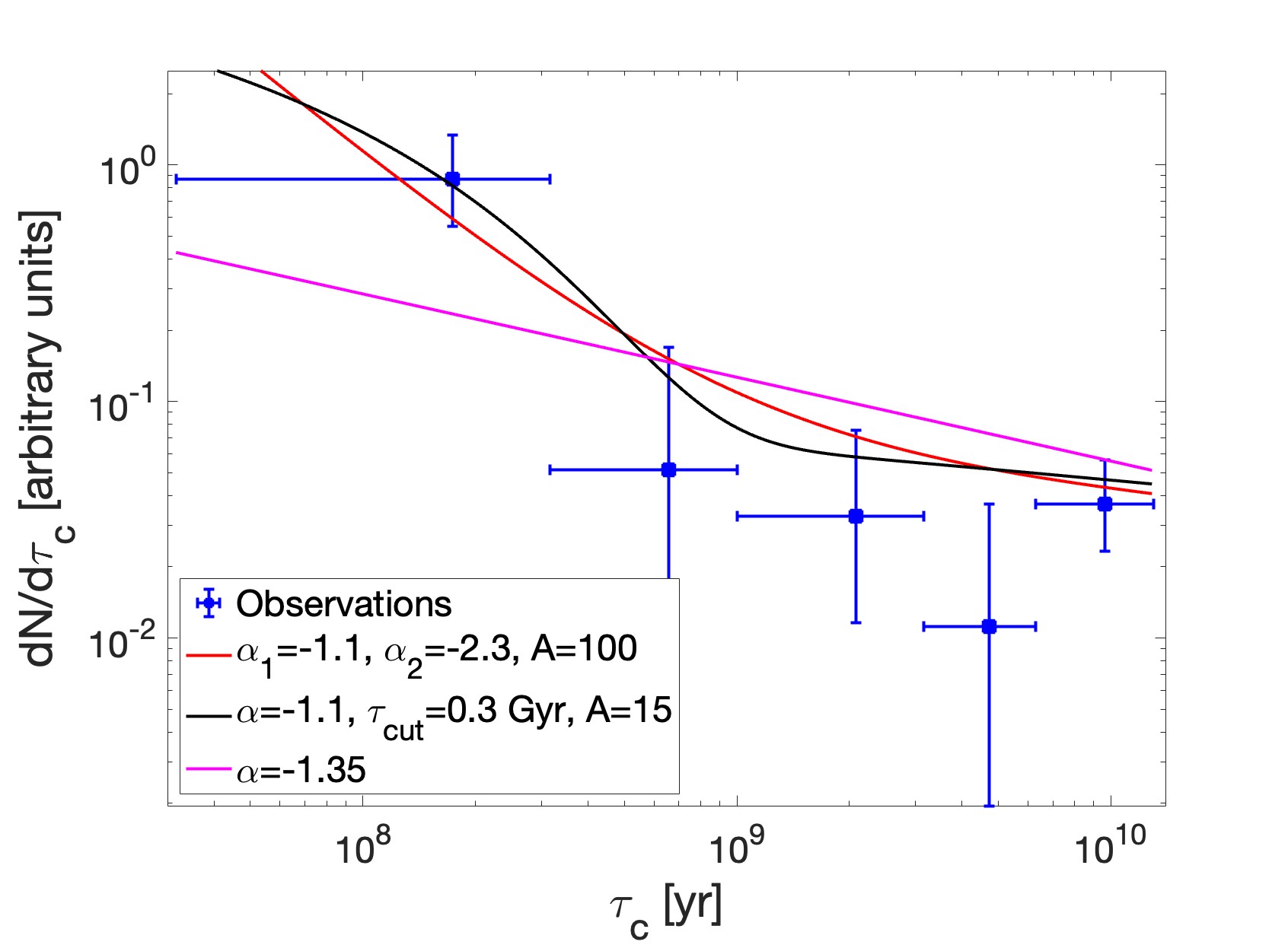}
    \caption{Observed distribution (crosses) of pulsar ages $\tau_c$---same as in Fig.~\ref{fig:dN_dtauc}---but showing only  the 18 rMSPs in DNS systems (Table~\ref{Table1}). Curves show the predicted distributions (Eq.~\ref{eq.taucdist}), for several of the DTD models considered for DNS mergers, based on the sample's distribution of time until merger, $\tau_{\rm gw}$. The best single-power-law DTD fit to the full $\tau_{\rm gw}$ range (Fig.~\ref{fig:dN_dtaugw}), with power-law index $\alpha=-1.35$, does not explain the excess of DNS pulsars 
    in the youngest bin. In contrast, the $\tau_c$ distributions predicted based on the best-fit two-population DTDs used to fit the $\tau_{\rm gw}$ distribution in Fig.~\ref{fig:dN_dtaugw2PL}, do reproduce the excess of young systems in the $\tau_c$ distribution.}
    \label{fig:dN_dtauc_DNS}
\end{figure}

Fig.~\ref{fig:dN_dtauc_DNS} shows again the observed $\tau_c$ distribution, but of the DNS sample only, now compared to the model prediction of Eq.~\ref{eq.taucdist}.
We plot the predictions for the three models considered above when fitting the $\tau_{\rm gw}$ distribution: a single-population of DNSs, with a single power-law DTD; two populations, each with a (different) power-law DTD;
and two populations, but one with power-law DTD with a cutoff. For each type of model, we show one with parameters within the 1$\sigma$ range of the best fit to the $\tau_{\rm gw}$ distribution, that also provides the best possible\footnote{Not surprisingly, not all the models that are favored by the fit to $\tau_{\rm gw}$  also provide a reasonable fit to $\tau_c$. Here (and in Figs.~\ref{fig:dN_dtaugw}-\ref{fig:dN_dtaugw2PL}) we choose models with parameters within the 1$\sigma$ range of the fit to $\tau_{\rm gw}$, that also provide a good fit to $\tau_c$, in the case of the two-population models, and the best (yet poor) fit in the case of the single power-law model.} description of the $\tau_{\rm c}$ distribution. The two-component models simultaneously fit well both $\tau_{\rm gw}$ and $\tau_c$. A single-power-law model, on the other hand, provides a poor fit to $\tau_{\rm gw}$ and an even worse fit to $\tau_c$.
We can now understand the excess of young DNS systems in the $\tau_c$ age distribution, noted in Fig~\ref{fig:dN_dtauc}. In effect, it is actually the result of a deficit, among all the old systems that have formed in the past, of the dominant numbers of small-separation, i.e. short $\tau_{\rm gw}$, systems, which 
have already merged. Naturally, this deficit is smaller or absent in models with DTDs that are more skewed to 
short $\tau_{\rm gw}$, via steeper power laws or cutoffs.

 In principle, we could constrain the DTD by fitting the above models to the 
 $dN/d\tau_c$ distribution, as we previously fit the models to the $dN/d\tau_{\rm gw}$ distribution, or even fit the DTD models jointly to the two observed distributions. However, the uncertainty in the current age of the systems (as estimated by $\tau_c$) is much larger than the uncertainty in the predicted time until merger ($\tau_{\rm gw}$), and we therefore choose to fit the DTD based on $\tau_{\rm gw}$ alone. Nevertheless, the comparison of the models to $\tau_c$ provides an independent test of the model DTDs. The fact that the two-population model indicated by the $\tau_{\rm gw}$ data reproduces well also the observed excess of DNS systems with small $\tau_c$, strengthens two of our previous conclusions: that $\tau_c$ is a reasonable estimator of the system age; and that a single power-law DTD provides a poor fit to the observed data, while a two-population model, dominated by short-lived systems, fits it well.

\cite{BeniaminiPiran2019} already noted an excess, among DNSs, of systems with $\tau_{\rm gw}<1$~Gyr, above a $\sim\tau^{-1}_{\rm gw}$ functional form that describes well the distribution at larger delays. They found a relatively mild excess, wherein the short delay population is not much larger than the $\sim\tau^{-1}_{\rm gw}$ population, while we find an excess that is much larger. The origin of this difference is that, 
contrary to us, they assumed that rMSPs lifetimes are short compared to the other timescales in the problem, in which case the $\tau_{\rm gw}$ distribution would
simply reflect the  DTD of DNSs. We have presented here evidence, however, that rMSP lifetimes are long, in which case the observed $\tau_{\rm gw}$ distribution is not simply the DTD, but rather it is described by Eq.~\ref{eq.taugw}. Furthermore,
in our DNS sample, which is by now larger than the one considered by \cite{BeniaminiPiran2019}, there is no longer an excess in the observed $dN/d\tau_{\rm gw}$ at short $\tau_{\rm gw}$ above $\tau^{-1}_{\rm gw}$, and it is in fact well-described by $dN/d\tau_{\rm gw}\propto\tau^{-1}_{\rm gw}$. However, we have shown that this is most likely an outcome of the superposition of two populations, each with a different $dN/d\tau_{\rm gw}$ distribution. 

\cite{Zevin2022} have recently estimated the DTD of short GRBs by analyzing the SFRs of GRB host galaxies, based on their broad-band photometry. Their best-fit power-law model for the DTD has index $\alpha=-1.8\pm0.4$ (90\% confidence interval), and $\alpha<-1.3$ at 99\% confidence, remarkably similar to our result for the fast/steep component of the DNS population, which would indeed dominate, by numbers, most of the observed merger events. \cite{Zevin2022} also obtain a constraint on the initial delay of the DTD, $t_i>72$~Myr at 99\% confidence.
We suspect, however, that this latter result may not be reliable, as the broad-band photometry they use permits estimating only simplified two-parameter SFHs for each galaxy, which may not afford the time resolution to resolve the DTD on such short timescales. 

A related issue is that of the host galaxy of GW170817, a passive nearby early-type galaxy (NGC~4993) with mean stellar population age $\sim$10 Gyr \citep{Nugent2022}. For a two-population DTD of the kind we have deduced here for DNS mergers, only $\sim 10^{-3}-10^{-4}$ of mergers occur between 10~Gyr and a Hubble time.  
The DNS merger rate today in the local universe is the convolution of the cosmic star-formation history with the DNS-merger DTD. Considering that the cosmic SFR $\sim 10$~Gyr ago ($z\sim2$) was an order of magnitude larger than today,
the fraction among all mergers that occur in local galaxies, having delays $>10$~Gyr, is expected to be correspondingly larger, but still only $\sim 0.1-1\%$. However,
\cite{Levan2017} have performed integral field spectroscopy near the site of GW170817, and deduce the presence of a significant 1~Gyr-old stellar population there, possibly related to a galactic merger event. Furthermore, among the ten lowest-redshift ($z<0.3$, lookback-time $<3.5$~Gyr) sGRB host galaxies in the sample of \cite{Nugent2022} (used for the analysis of \citealt{Zevin2022}), two have a stellar population mass-weighted age (time between star formation and the time of the sGRB) $t_{\rm m}\sim8-9$~Gyr, quite similar to GW170817.
Local sGRB hosts similar to NGC~4933 are therefore not unusual, and perhaps 
such more-recent star-formation episodes in old quiescent galaxies are not uncommon. Finally, it is also important to remember that our DTD is estimated based on a Milky-Way DNS sample, while the actual DNS-merger DTD elesewhere could depend on cosmic time and environment.

\section{How to make knees}\label{sec:knee}
\subsection{Chemical evolution calculation}
To investigate how a "prompt" DTD for DNS mergers and their ensuing kilonovae, as we have deduced in \S\ref{sec:pulsars}, above, may be relevant for chemical evolution and the [Eu/Fe] "knee problem" (see \S\ref{sec:intro}), we follow \citet{MaozGraur2017}
and, as a proof of concept, perform a simple chemical evolution calculation that predicts the evolution of stellar generations in the [X/Fe] vs. [Fe/H] plane with minimal assumptions and essentially no free parameters. 
Our calculation approximates the thick-disk component of the Galaxy as a single-zone closed box of stars and gas with instant full mixing, and is equivalent to chemical evolution models with no losses of metals due to outflows, no dilution by pristine gas from inflows, and gas depletion time equal to the star-formation time. While likely a highly simplified version of reality, this calculation may nevertheless capture the essence of the chemical-evolution process. As we show in \S\ref{sec:MW_iron}, at least the approximation of no significant losses of metals from the system is empirically justified. 

We use actual measurements of Milky Way's thick-disk stellar population to represent the star-formation history, ${\rm SFR}(t) $. The CC-SN rate tracks the star-formation rate, $R_{\rm cc}(t)=(N_{\rm cc}/M_*){\rm SFR}(t) $, scaled by the number of core-collapsing stars per unit formed stellar mass, $N_{\rm cc}/M_*$. For a standard \cite{Kroupa2001} initial mass function (IMF), and assuming
all stars more massive than $8M_\odot$ explode,  $N_{\rm cc}/M*=0.01$. It has been argued (e.g., \citealt{Smartt2015,Kochanek2020}) that some fraction of massive stars collapse directly to black holes, without a SN explosion, which would lower  $N_{\rm cc}/M_*$ by some fraction. However, \citet{Strotjohann2023} have recently shown that, to date, there is no statistically significant evidence for this so-called "red-supergiant problem", and we therefor ignore such a fraction. 

The SN Ia rate, $R_{\rm Ia}$, is the 
convolution of the SFR with the SN Ia DTD, $R_{\rm Ia}={\rm SFR}(t)\circledtimes D_{\rm Ia}(t)$ .
For the DTD of SNe Ia, we assume the observationally determined form \citep{MaozGraur2017,FreundlichMaoz2021}, with an initial delay $t_i=40$~Myr, followed
by an abrupt rise to a falling power law, $t^{-1.1}$. $D_{\rm Ia}(t)$ is normalized, when integrated over 13.7 Gyr, to  either $N_{\rm Ia}/M_*=(0.0013\pm0.0001) M_\odot^{-1}$ SNe Ia per unit formed stellar mass, as measured in field galaxies \citep{MaozGraur2017}; or, $N_{\rm Ia}/M_*=(0.0031\pm0.0011) M_\odot^{-1}$, which is the current estimate for the SN Ia DTD normalization in early-type galaxies in clusters and in the field \citep{FreundlichMaoz2021,Toy2023}. The latest empirical estimate of mean iron mass yield per CC-SN, averaged and weighted over the various main CC-SN types, is $\bar y_{\rm Fe,cc}=(0.058\pm0.007) M_\odot$ \citep{Rodriguez2023}. For the mean iron yield of SNe Ia we take  $\bar y_{\rm Fe,Ia}=(0.7\pm 0.05) M_\odot$ \citep{Howell2009}. Let us denote with $(\alpha/\rm Fe)_{\rm cc} $ the mean yield ratio between an $\alpha$ element (or elements) and iron in CC-SNe. However, we do
not require to know the value of this parameter, as it cancels out in the abundance ratios we consider, which are always relative to solar abundance ratios. The iron masses originating from CC-SNe and from SNe Ia, respectively, and accumulated in stars and in the ISM at time $t$ after initiation of star formation are
\begin{equation}
\label{EQFecc}
    M_{\rm Fe,cc}(t)=\int_0^{t}  R_{\rm cc}(t') ~  \bar y_{\rm Fe,cc} ~ dt'   ,
\end{equation}
\begin{equation}
\label{EQFeIa}
    M_{\rm Fe,Ia}(t)=\int_0^{t}  R_{\rm Ia}(t') ~  \bar y_{\rm Fe,Ia} ~ dt'   .
\end{equation}
The accumulated mass of an $\alpha$-element is
\begin{equation}
\label{EQalphamass}
M_{\rm \alpha,cc}(t)=\int_0^{t}  R_{\rm cc}(t') ~  \bar y_{\rm Fe,cc} ~(\alpha/{\rm Fe})_{\rm cc}~ dt'  ,
\end{equation}
where we have assumed, as common, that $\alpha$-elements are produced predominantly by CC-SNe. We can now calculate [$\alpha$/Fe]$(t)$ and[Fe/H]$(t)$ as
\begin{equation}
\label{EQalphaFe}
\left[\frac{\alpha}{\rm Fe}\right]=\log\frac{M_{\rm \alpha,cc}(t)/[M_{\rm Fe,cc}(t) +  M_{\rm Fe,Ia}(t)]}{M_{\rm \alpha,cc}(t_\odot)/[ M_{\rm Fe,cc}(t_\odot) +  M_{\rm Fe,Ia}(t_\odot)]},
\end{equation}
and 
\begin{equation}
\left[\frac{\rm Fe}{\rm H}\right]=\log\frac{ M_{\rm Fe,cc}(t) +  M_{\rm Fe,Ia}(t)}{M_{\rm Fe,cc}(t_\odot) +  M_{\rm Fe,Ia}(t_\odot)},
\end{equation}
where $t_\odot$ (the only free parameter in the calculation) is the time at which solar abundance is attained.

\subsection{The {\rm [$\alpha$/Fe]} knee}

\begin{figure}
    \centering
    \includegraphics[width=9.5cm]{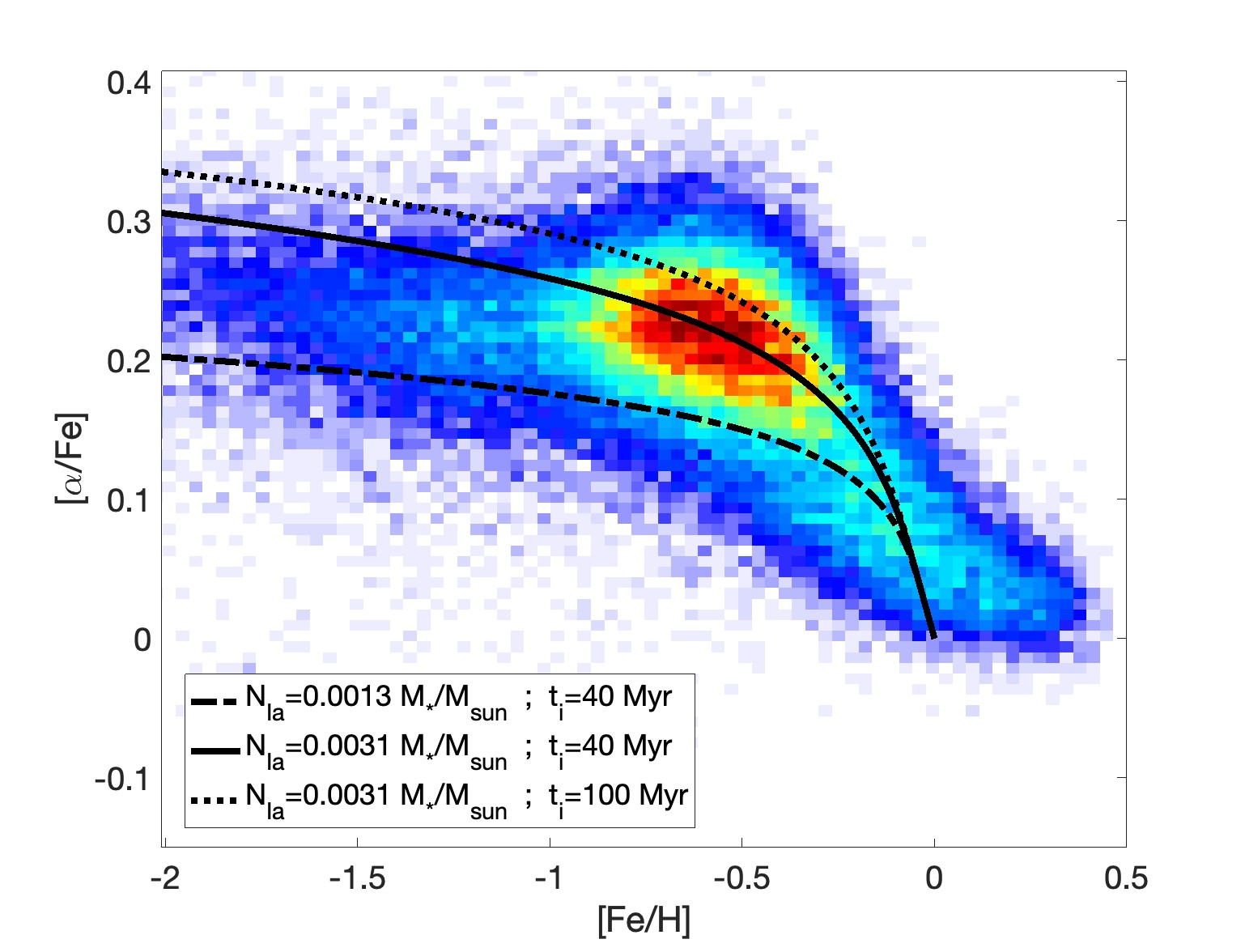}
    \caption{Color map representing the number density of $\sim 100,000$ subgiant stars from \cite{XiangRix2022} belonging primarily to the thick-disk population of the Milky Way, in the abundance-ratio plane [$\alpha$/Fe] vs. [Fe/H] (highest densities are marked red, lowest blue). Curves are results of our all-empirical-based chemical evolution calculation, for two possible values of the normalization and two values of the initial delay time, $t_i$, of the $D_{\rm Ia}\sim t^{-1}$ SN Ia DTD..
    The observed abundance distribution is well-reproduced  by our calculation,
    as already concluded by \cite{MaozGraur2017}. The initial delay has a weak effect on the predicted curves.}
    \label{figalphaRix}
\end{figure}
We begin by testing the ability of this simple chemical evolution model to reproduce the well-known "high-$\alpha$" sequence of
thick-disk stars in the [$\alpha$/Fe] vs. [Fe/H] plane\footnote{All abundance measurements we cite here and forthwith are relative to solar abundances from \cite{Asplund2009}}. Fig.\ref{figalphaRix} shows, in this parameter plane, the stellar number density of stars from \citet{XiangRix2022}
having azimuthal action $J_\phi<1500$~kpc~km~s$^{-1}$ (i.e. members of the thick-disk and halo populations).
The over-plotted curves are the results of our chemical-evolution calculation, using for ${\rm SFR}(t)$ the empirically determined thick-disk star-formation history of \cite{Fantin2019} based on analysis of WD numbers and ages. The \cite{Fantin2019} SFH is a skewed Gaussian peaking $\sim 10$~Gyr ago with a full-width at half maximum of about 1~Gyr. We do not use
the more-slowly evolving thick-disk SFH of \citet{XiangRix2022}, as we suspect it suffers from contamination by other Galactic populations and by systematic errors, as evidenced, e.g., by many of the stars having ages much older than the Universe.  
Nevertheless, the data of \citet{XiangRix2022} concur that the thick disk was formed in a burst of star formation that peaked sharply 
$\sim 10$ Gyrs ago.  Moreover, their data show that the typical stars that formed at the time of the SFR peak had [Fe/H]$\approx-0.5$ and   [$\alpha$/Fe]$\approx 0.25$, precisely the values at the "knee" in the thick-disk sequence. The drop in [$\alpha$/Fe] coincidess with the sharp drop in the star formation rate, as predicted by our model and, barring an unlikley coincidence, seems to have no relation to the initial delay, $t_i$, in the DTD of type Ia SNe.

Two of the curves in  Fig.\ref{figalphaRix} show our calculation for the two possible values of the normalization, $N_{\rm Ia}/M_*$, of the SN Ia DTD,  whether in field galaxies \citep{MaozGraur2017} or in early-type galaxies\citep{FriedmannMaoz2018,FreundlichMaoz2021,Toy2023} with the usual $t_i$ value of 40 Myr, which recovers the conclusion of \cite{MaozGraur2017} that the high-normalized early-type-environment DTD of SNe Ia  best reproduces the observed high-$\alpha$ sequence of the thick-disk stellar population. This may be a reasonable choice, given the possible similarity between conditions during star formation in the very-young Milky Way and in early type galaxies.
We note, however, that any change of parameters that sufficiently raises the contribution, relative to that of CC-SNe, of SNe Ia to the total accumulated iron, will have the same effect, for example, a smaller mean iron yield of CC-SNe, or a fraction less than unity of massive stars that explode, rather than collapsing without a SN explosion. Furthermore, a more mundane explanation is that $\alpha$-element abundances have systematic uncertainties of roughly $\pm 0.15$~dex (\citealt{Weinberg2023}, see further below), which could easily bring to accord the data and the model that has the standard SN~Ia normalization of the DTD. The principal success of the model which we point out is the reproduction 
of the general structure of a sloping plateau plus a knee at roughly the right location in the diagram.
We re-emphasize that all of the input parameters to our model are measured from observations; other than $t_\odot$, our calculation has no free parameters. 

Our calculated evolution track demonstrates that the knee is a direct consequence of the sharp decline, $\sim10$~Gyr ago, of the SFR and of its closely tracking, $\alpha$-element-producing, CC-SNe. The knee has little to do with the initial SN Ia delay $t_i$. To illustrate this point, Fig.\ref{figalphaRix} shows also our abundance evolution curves for two values of $t_i$: 40~Myr, which is often assumed, being the time required for stellar evolution to produce the first, most massive, WDs; and 100~Myr (also often used in chemical evolution models), which is an initial delay value that would already be in tension with SN Ia DTD measurements, e.g. \cite{MaozMannucciBrandt2012}. These two extremes for the value of $t_i$ have only a weak influence on the location or shape of the knee. Rather than resulting, as often claimed, from a rise in SN Ia rate and iron production, the knee is the result of a decline in the rate of SNe Ia, and the iron that they produce, yet a decline that is milder than that of the CC-SNe. This confirms the explanation of the knee proposed by \cite{MaozGraur2017}, based on a hypothesized SFR that reproduced the abundance-ratio measurements (a hypothesized SFR that matches closely the empirical one we use here). As noted above, this result is supported by the data of \cite{XiangRix2022} and by the theoretical study by \cite{Mason2023},  who found, using detailed galaxy formation simulations and chemical evolution modeling, that abundance-diagram knees in realistic galaxies are always the result of SFR declines, and not of delayed SN Ia rate "kick-ins".

\subsection{The {\rm [Eu/Fe]} knee}
Having established that our chemical evolution calculation, with its all-empirical input parameters, reproduces the "high-$\alpha$" sequence of
thick-disk stars in the [$\alpha$/Fe] vs. [Fe/H] plane, we now turn to the r-process elements, and attempt to reproduce their observed abundance sequence, informed by our findings in \S\ref{sec:pulsars} regarding the DTD of DNS mergers.  
To calculate, for instance, Eu-abundance evolution, we replace the core-collapse SN rate from the previous calculation with a DNS-merger-driven kilonova rate, given 
by the convolution of the SFR with a DTD for kilonovae,
$R_{\rm kn}={\rm SFR}(t)\circledtimes D_{\rm kn}(t)$.
As already discussed in \S\ref{sec:pulsars} and, in more detail, in Appendix~B, $D_{\rm kn}(t)$=0 at low $t$, up to an initial delay $t_i$, dictated by the stellar and binary evolution parameters of the merging binary systems. The single-power-law DTD models we have considered rise abruptly from zero to a maximum at $t_i$ and then 
fall off as $t^{\alpha}$. The two-DNS-population models considered in \S\ref{sec:pulsars} have a DTD that is either the sum of two 
single-power-laws with indices $\alpha_1$ and $\alpha_2$, and normalization ratio $A$, or the sum of a power-law with another 
power law with an exponential cutoff, $\tau^{-1}\exp(-\tau/\tau_{\rm cut})$, again with a number ratio $A$ between the populations. For the purpose
of our chemical evolution calculation, $t_i$ represents the time elapsed between the explosions of the bulk of the iron-producing core-collapse SNe from a stellar generation,
and the first kilonovae from DNS mergers from that stellar generation (see Appendix B). Therefore, 
$t_i$ could be close to zero \citep[e.g.,][]{Beniamini2024}, or it could be somewhat longer, e.g. due to additional binary-evolution timescales
associated with DNSs, compared to isolated NSs from SN explosions. We explore a range of reasonable 
values for $t_i$, from 1 to 10 Myr.
   
In analogy to Eq. \ref{EQalphamass}, the accumulated Eu mass from kilonovae is
\begin{equation}
M_{\rm Eu,kn}(t)=\int_0^{t}  R_{\rm kn}(t') ~  \bar y_{\rm Eu} ~ dt'  ,
\end{equation}
where $\bar y_{\rm Eu}$ is the mean Eu yield per kilonova explosion, which we do not need to know for our calculation, as the end result is normalized to solar abundance ratios. We then have
\begin{equation}
\left[\frac{\rm Eu}{\rm Fe}\right]=\log\frac{M_{\rm Eu,kn}(t)/[M_{\rm Fe,cc}(t) +  M_{\rm Fe,Ia}(t)]}{M_{\rm Eu,kn}(t_\odot)/[ M_{\rm Fe,cc}(t_\odot) +  M_{\rm Fe,Ia}(t_\odot)]}.
\end{equation}

\begin{figure}
    \centering
    \includegraphics[width=9.5cm]{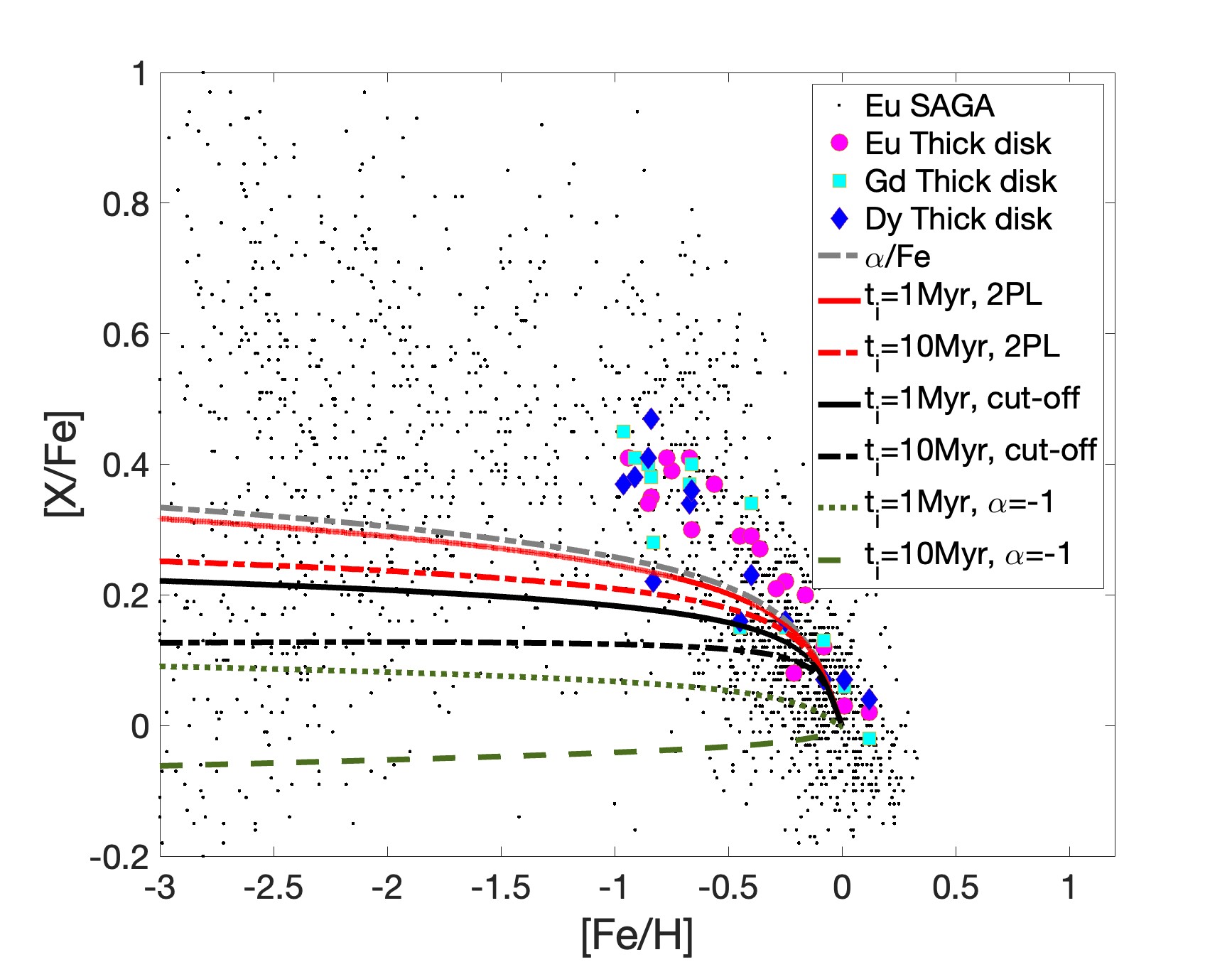}
    \caption{Observed r-process-element abundance ratios, compared to our chemical-evolution calculation. Eu abundances ratios (small dots) are compiled from the SAGA database \citep{Suda2008}. Large symbols are measurements in thick-disk stars
of Eu \citep{Mishenina2013}, gandolinium and dysprosium \citep{Mishenina2022}. Curves show our chemical evolution calculations for kilonovae with the various DTDs considered in \S\ref{sec:pulsars}: A $t^{-1}$ DTD (lowest two dotted and dashed curves) which, as expected, does not produce a "plateau-plus-knee" structure; and four different two-population kilonova DTD models based on the DNS-sample analysis in \S\ref{sec:pulsars}. 
Two of these four models have DTDs that are combinations of two power laws, and two of them are combinations of a power law with another power law that has a cutoff. Each model shown has one of two extreme values of initial delay $t_i$ that we consider. All of these two-population models reproduce a plateau-plus-knee structure in the diagram. Also shown is the model [$\alpha$/Fe] curve reproducing the data in Fig.~\ref{figalphaRix}. The observed r-process-element plateau appears systematically higher by $\sim 0.1$~dex than the observed $\alpha$-element plateau, and higher than all the r-process models, but systematic uncertainties in abundance-ratio estimates are significantly larger than this offset, both for $\alpha$ elements and for r-process elements (see text).}
    \label{figEu}
\end{figure}

 Fig.~\ref{figEu} shows measured Eu stellar abundances as compiled from the Stellar Abundances for Galactic Archaeology (SAGA)  database \citep{Suda2008}.
A shortcoming of these data, for the current application, is that they are not labeled according their Galactic kinematic component, and thus presumably could include thin-disk,
thick-disk, and halo stars, blurring the possibly different evolution locii and their salient features. Furthermore, this large compilation is heterogeneous and difficult to separate according to precision and accuracy. \citet{Mishenina2013} have published Eu abundances for a sample of 18 strictly thick-disk stars. For most of these stars, \cite{Mishenina2022} have further reported abundances of the elements gandolinium (Gd) and dysprosium (Dy), 
which like Eu, are also predominantly produced through the r-process, and may therefore also come from kilonovae and follow the same abundance patterns  (all three elements are neighbors in the periodic table and are likely formed at the same sites). We show in Fig.~\ref{figEu} the abundance ratios for all three r-process elements for these thick-disk stars. The r-process-element measurements in Fig.~\ref{figEu} illustrate the morphology of this abundance sequence, with a plateau-plus-knee structure similar to that of  $\alpha$-elements. This, even though kilonovae have traditionally been expected to have a DTD with a broad $t^{-1}$ form similar to  SNe Ia, rather than tracing the rate of CC-SNe like  $\alpha$-elements (see \S~\ref{sec:intro}).

We overplot in Fig.~\ref{figEu} curves with the outputs of our simple chemical evolution calculation for the r-process elements, using for the DTD of kilonovae several of the single-population and two-population DNS models that we have constrained in \S\ref{sec:pulsars}, based on the observed Galactic DNS population.
The overall structure seen in the diagram, particularly the plateau and the knee, can be reproduced with the two-population DTDs that we have assumed. Either the steep power-law component ($\sim t^{-2}$), in the two-power-law models, or the exponential-cutoff component, in the cutoff models, can serve to limit the time, after star formation and core-collapse SNe, during which r-process elements are produced, as compared to the extended $\sim t^{-1}$ form of the DTD of SNe Ia.
 We further show in Fig.~\ref{figEu}
 (dotted curve) that a single-power-law DTD of the form $\sim t^{-1}$ for kilonovae fails to produce the plateau-plus-knee structure, irrespective of $t_i$ (as has been previously known, see \S~\ref{sec:intro}).

Fig.~\ref{figEu} shows also the model [$\alpha$/Fe] curve that reproduces the data in Fig.~\ref{figalphaRix}.
It is important to realize that the quantity which [$\alpha$/Fe] actually measures, at a given cosmic time, or at a given metallicity [Fe/H], is the fraction of the total accumulated iron mass in a star that was created by CC-SNe by that time, relative to that fraction in the Sun (see Eqns.~\ref{EQalphamass}-\ref{EQalphaFe}). Indeed, it is easy to see that, at a very early time in the SFH, before any material enriched by SNe~Ia has been incorporated into a star (i.e. at $t<t_i$ for SNe Ia), the fraction of the iron from CC-SNe is one, and therefore [$\alpha$/Fe]$^{-1}$ at that time is simply the fraction of iron in the sun contributed by CC-SNe. 
This is the reason that all of the different $\alpha$-elements that are produced predominantly by CC-SNe (e.g. O, Mg, Ca) have nearly identical loci in the [$\alpha$/Fe] vs. [Fe/H] plane (even though their absolute abundances relative to Fe, when not normalized to solar ratios, vary greatly). Similarly, if Eu were produced promptly in the explosions of massive stars that 
closely tracked the SFR, one would expect [Eu/Fe] to behave identically to [$\alpha$/Fe]. In our model for the kilonova DTD,
where kilonovae are somewhat delayed with respect to CC-SNe, the [Eu/Fe] evolution curve is always {\it below} the [$\alpha$/Fe] curve, as the kilonovae need to "catch up" with the core-collapse SNe and the iron that they have already contributed at any given time.

Examining Fig.~\ref{figEu} in this context, it is interesting to note that, in the range starting at [Fe/H]$>-1$, for which [Eu/Fe] data exist specifically for thick disk stars, and up to the knee, the observed [Eu/Fe] values appear to be (with a large scatter) {\it higher} than [$\alpha$/Fe]
by $\sim 0.05-0.15$~dex. An offset above the $\alpha$-elements of the thick-disk r-process points continues also toward [Fe/H]=0 and beyond. An [Eu/Fe] ratio higher than [$\alpha$/Fe]  at a given [Fe/H] is something that our simple model cannot reproduce, as already noted above, unless we introduce a time- or metallicity-dependent Eu production in a way that will mimic the $\alpha$-elements locus. This limitation is true for all Galactic chemical evolution models, for the reasons noted above. 
However, the $\sim 10-40\%$ excess in Eu abundance is quite possibly  a systematic error in the measurements and/or in the conversion of Eu absorption-line equivalent widths to abundances, by means of models for low-metallicity stellar-atmospheres. \cite{Weinberg2023}, for instance, note  that even $\alpha$-element abundances suffer from systematic uncertainties of $\pm0.15$~dex, as evidenced by the plateau offsets among different $\alpha$ elements.
For r-process elements, the uncertainties could well be larger. For example, \cite{Mishenina2013} estimate typical systematic uncertainties in their own measured Eu abundances of  $\pm0.10$~dex. \cite{Hinkel2016} compared the element abundance estimates derived autonomously by six different research groups from the same spectral data for four stars. For Eu, the estimates of the groups differed over a range of 0.15 to 0.47 dex for the four stars, larger than the offsets
between the Eu plateau and the various model curves in  Fig.~\ref{figEu}.
Given the current accuracy of r-process abundance estimates, we do not consider it presently warranted to add to our calculation any complexity  that might explain the offsets, for instance, a dependence of r-process yields on metallicity.

\section{Discussion}\label{sec:discussion}

We have shown that the general characteristics of the locus of thick disk stars in the [Eu/Fe] vs [Fe/H] diagram, particularly the knee at [Fe/H]~$\sim -0.5$, can be reproduced by a simple chemical evolution model, based only on empirical input for the thick-disk SFH and for SN Ia and CC-SN  parameters, but with a DTD for kilonovae, dominated by short-lived NS binaries, having a steep, $\sim t^{-2}$, power-law form or, alternatively, $\sim t^{-1}$ with an exponential cutoff with characteristic time $\tau_{\rm cut}\approx 300$~Myr. At delays $\gtrsim 1$ Gyr, most of the DNS mergers are due to a sub-dominant population with DTD $\sim t^{-1}$. The case for the existence of such a kilonova DTD  is motivated by our analysis, in \S\ref{sec:pulsars}, of the observed distributions of spin-down time, $\tau_c$, and time until merger, $\tau_{\rm gw}$, of the known systems of rMSPs in DNSs. 

This DTD for DNS mergers and kilonovae suggests that there may be more than one formation channel for DNS systems in the field (in addition to a formation channel in globular clusters). In the dominant channel, the formation of  bound DNSs, for whatever physical drivers behind it, prefers short periods. The DTD implies a distribution of the initial DNS separations, $a$, that is either very steep, $dN/da \propto a^{-k}$ where $k \approx 4-6$, or that there is a cut-off in the initial separation distribution at $a \sim 2R_\odot$ (corresponding to a merger time of $\sim300-500$~Myr for two NSs). The sub-dominant channel, on the other hand, has a roughly constant distribution per logarithmic interval of separation, at least up to a separation of $\sim 50~R_\odot$ (corresponding to a merger time of $\sim 10^5$~Gyr).  


A number of previous works have proposed that there are two, or even  more, different formation channels for DNSs, based on the observed properties of Galactic DNSs \citep[e.g.,][]{Piran2005,DallOsso2014,Beniamini2016,Chruslinska2017,Andrews2019}. 
We speculate that the strong preference for an initial DNS separation of $\lesssim 2R_\odot$ could be the outcome of some physical requirement on the final binary-evolution phase of the system before DNS formation. For example, such a small separation could be required for (or be the result of) the second pre-SN star
undergoing enough mass stripping or depletion by the first NS, such that its explosion is weak, with a low kick that leaves the system bound (as in the scenario of \citealt{Beniamini2016}; see also \citealt{Nugent2024} for recent evidence of small NS kicks in DNS formation in dwarf galaxies). In any event,
our finding here that the large majority of DNSs that merge within a Hubble time have a distribution of separations that are strongly biased toward small ones, with correspondingly short times until merger, may provide one more  clue toward understanding the formation of bound DNS systems, the explosion of some them as kilonovae, and their role in cosmic element production. 


\section{Epilogue: the Milky Way's iron budget} \label{sec:MW_iron}
Our simple chemical evolution calculation, above, which reproduces remarkably well the observed plateau-plus-knee sequences of
$\alpha$- and r-process elements, may be criticized on the grounds that it is oversimplified. In particular, our calculation assumes that the Galaxy is a "closed box", in which all explosive nucleosynthesis products have been retained, whereas metal-depleting galactic winds are thought to actually be important to the chemical evolution budget. 

To address the closed-box question, we compare here, using the same empirical parameters we have used for our chemical evolution calculation, the total mass of iron residing today in the Galaxy to the total iron mass produced by SNe in the course of the Galaxy's lifetime. The Galaxy consists of the following baryonic components. The stars, with total mass $M_*$, have a mean iron abundance $\bar Z_{*, \rm Fe}\approx(0.75\pm 0.25) Z_{\odot,\rm Fe}$ of the solar iron mass abundance, $Z_{\odot,\rm Fe}=0.00137$ \citep{Asplund2009}. This value and uncertainty range for the mean abundance of stars can be seen in \cite{Buder2021},
who compared the stellar metallicity distributions in several large surveys, and found systematic differences within this range, resulting from the differing Galactic regions probed and the different sample selection criteria of those surveys. The cold interstellar medium (ISM) constitutes a
  fraction $f_{\rm ism}\approx 0.22\pm 0.02$ of the stellar mass \citep{McMillan2017}, and has a metal abundance of non-refractory elements that is solar to within a few per cent \citep{Ritchey2023}. However, at least 90\% of the iron in the ISM is depleted onto dust grains \citep{Psaradaki2023}. We can therefore safely account for the total iron mass in the ISM (whether in the gas or within dust particles) as 
  $M_* f_{\rm ism} Z_{\odot,\rm Fe}$.
  The circumgalactic medium has a mass comparable to the stellar mass, $f_{\rm cgm}\approx 1.0\pm 0.2$, and a mean abundance $\bar Z_{\rm cgm, Fe}\approx (0.5\pm0.1) Z_{\odot, \rm Fe}$ \citep{Tumlinson2017,FaermanSternbergMcKee2020}.
  The total iron mass today in the Milky Way is thus,
  \begin{equation}
  M_{\rm Fe,0}=M_*\left(\frac{\bar Z_{*,\rm Fe}}{Z_{\odot,{\rm Fe}}}+f_{\rm ism}+f_{\rm cgm}\frac{\bar Z_{\rm cgm, Fe}}{Z_{\odot,\rm Fe}}\right)Z_{\odot,\rm Fe}\approx0.0020M_*  ~ .
  \end{equation}
The iron mass produced by SNe that have exploded over a Hubble time is proportional to the formed stellar mass. For any not-extremely young stellar population, the formed stellar mass is $M_*/(1-r)$, where $r=0.4$  (for a \citealt{Kroupa2001} IMF, as consistently assumed throughout) is the "recycling factor", i.e. the fraction of the mass from a newly formed stellar population that is lost to stellar winds and SN ejecta. The total iron mass formed over the Galaxy's history is then,   
 \begin{equation}
  M_{\rm Fe, formed}=\frac{M_*}{1-r}\left(\frac{N_{\rm cc}}{M_*}\bar y_{\rm Fe, cc} + \frac{N_{\rm Ia}}{M_*}\bar y_{\rm Fe, Ia}\right)\approx0.0025M_* ~ ,
  \end{equation}
  where the SN numbers per formed stellar mass and the mean iron yields per SN are as previously defined.
  The fraction of the iron mass produced that is still bound in the Galaxy is therefore
  \begin{equation}
  \frac{M_{\rm Fe,0}}{M_{\rm Fe, formed}}=0.81\pm0.10 ~,
  \end{equation}
  where the uncertainty is the root-mean-square of a Monte-Carlo calculation where each of the input parameters is allowed to vary randomly with a flat probability over its systematic uncertainty range, noted above.
 The result could be even higher if, e.g., some fraction of massive stars collapse without a SN explosion; or somewhat lower if, as we have seen some evidence here, the Hubble-time-integrated number of SNe Ia per formed stellar mass of thick-disk stars (which constitute some fraction of the Galaxy's stellar mass) is higher than for other populations.

 In any event, it is  hard to avoid the conclusion that the Milky Way has retained $\sim70-90\%$, and possibly all, of the metals that have been created within it. 
 About 50\% of those metals are currently locked within stars, 15\% are in the interstellar medium, and 35\% are in the circumgalactic medium.
 Thus, even if the Galaxy is not a truly closed box, it is a fairly well-sealed one. 
 No less remarkable is the fact that the current and the formed iron masses, estimated in completely unrelated measurements and environments (abundances in Galactic stellar atmospheres and interstellar gas, versus SN rates and iron yields in both nearby and distant galaxies), turn out to be so close to each other. This suggests that we have not missed any major iron-producing population, and that we may have a good quantitative understanding of the processes involved in iron making.   

\section*{Acknowledgements}
We thank Vicky Kaspi, Amir Levinson, and Paz Beniamini for useful discussions and comments.
This paper is part of a project that has received funding from the European Research Council (ERC) under the European Union’s Seventh Framework Programme, grant agreement No. 833031 (PI: Dan Maoz) and grant agreement No. 818899 (PI: Ehud Nakar).

\section*{Appendix A\\ Possible selection effects in the time-till-merger $\tau_{\rm gw}$ distribution}\label{App:selectioneffects}
In order to safely use the observed $\tau_{\rm gw}$ distribution to constrain the DNS merger DTD, 
we need to understand if there are any selection effects in the discovery of pulsars, in general, and rMSP DNSs in particular, that are correlated with $\tau_{\rm gw}$. There certainly do exist pulsar selection effects related to the rMSP properties, such as the beam opening angle and the radio luminosity, but these do not appear to be correlated with $\tau_{\rm gw}$. Another type of selection effect may have to do with the location of a DNS in the Galactic disk. Younger systems are more likely to be closer to their birth location near the Galactic plane, although this is probably insignificant for systems with low center-of-mass velocity, as seems to be the case for quite a few DNSs \citep{Tauris2017}. Systems near the plane are harder to detect, and the sky coverage of the various pulsar surveys as a function of the height above the plane is not necessarily uniform. Finally, a well-known selection effect is that systems with high orbital acceleration are harder to detect due to Doppler smearing of a pulsar signal \citep[e.g.,][]{Pol2021}. This results in a reduced sensitivity to systems with small period and large eccentricity, i.e., to system with small values of $\tau_{\rm gw}$. These selection effects works in the direction opposite to the observed effects that we see, of an excess of short  $\tau_{\rm gw}$ and short $\tau_c$ systems. If these selection effects are significant in the detection of the DNS sample, then the excess we have noted is an underestimate of the true excess.

The combination of the above selection effects have been studied by various authors. For example, \cite{ColomIBernadich2023}, in order to estimate the total Galactic DNS merger rate,  have
simulated the detection, by each of the DNS surveys, of every known DNS that will merge within a Hubble time. For each such DNS, they estimate what is the fraction of a population of DNSs with similar properties that would be detected by the surveys that were carried out till now. They find that indeed, on average, DNSs with lower $\tau_{\rm gw}$ are harder to detect,  but not by much (see their table 4). Although we cannot rule out the possibility of unknown selection effects, it seems that the available sample of rMSP DNSs can provide a representative sample of the Galactic distribution of $\tau_{\rm gw}$, bearing in mind that DNSs with small values of $\tau_{\rm gw}$ might be somewhat under-represented in our sample (which of course only strengthens the case for an excess of DNSs with small $\tau_c$ and small $\tau_{\rm gw}$). A complete sample of $\tau_{\rm gw}$ could also serve to constrain the value of $t_i$, but we do not attempt this, as our current observed sample may suffer from selection effects that discriminate against DNSs with lifetimes $\lesssim 50$ Myr.  

\section*{Appendix B\\ The different types of "initial delays" in the DTD, where and how to use them}
As already noted in footnote~\ref{tifootnote}, there are a number of different "initial delays" that can be defined when considering the DTD of DNS mergers and, in principle, different variants of the initial delay need to be used in different aspects of the calculations in this paper. One initial delay is the time $t_{i, {\rm dns}}$, between star-formation and the formation of the first DNS systems, i.e. the formation of the two neutron stars in the binary following the SN explosions of their progenitor massive stars, where in the process, at least one of the NSs was spun up into a rMSP via accretion from its pre-NS companion. At time $t_{i, {\rm dns}}$, set by the stellar and binary evolution timescales of binary stars that end up as DNS rMSPs, the first DNS system begins its inspiral toward merger. A second type of initial delay is the time $t_{i, {\rm gw}}$ between DNS formation (as defined above) and the first DNS merger events. The earliest merger events are those in which the (small) separation and (high) eccentricity of the newly born DNSs lead to the shortest $\tau_{\rm gw}$. Finally, the sum of these two timescales, which we defined as $t_i$ in \S\ref{sec:pulsars}, is the times from star formation until the first DNS mergers.

When calculating the present-day distribution of times-till-merger $dN/d\tau_{\rm gw}$ (Eq.~\ref{eq.taugw}),
the $t_i$ implicit in the DTD in the integrand, $D(\tau_{\rm gw}, t)$, is indeed $t_i$ (the longest of the three types of initial delays defined above), as no mergers with initial delays smaller than this exist.
 At any given value of present-day observed $\tau_{\rm gw}$ (not necessarily short ones), there are contributions from 
the DNS populations that were formed in the past, from time $t_{i, {\rm dns}}$ ago to time $t_0$ ago, when star formation in the Galaxy began, and therefore the lower limit of the integral, should be $t_{i, {\rm dns}}$. When considering a power-law DTD (which arises as a result a power-law distribution of initial DNS separations), the DTD, at $\tau>t_i$, will have the form 
$D(\tau)\propto (\tau-t_{i,{\rm dns}})^\alpha$, and zero otherwise. Thus, changing variables $\tau-t_{i,{\rm dns}} \rightarrow \tau'$ we obtain Eq.~\ref{eq.taugw} with a correction to the integrand: ${\rm SFR}(t_0-t+t_{i, {\rm dns}})$; and to the upper limit of the integral: $t_0-t_{i, {\rm dns}}$. Given that  ${\rm SFR}(t)$ is roughly constant and $t_0 \gg t_{i, {\rm dns}}$, both of these corrections are negligible.


For chemical evolution calculations of the type in \S~\ref{sec:knee}, the relevant initial delay to use is $t_{i, {\rm gw}}$. The reason is that the main physical process that we attempt to understand
is the delay between the metal enrichment by core-collapse SNe and by kilonovae from DNS mergers (or from SNe~Ia). As core-collapse SNe, just like DNSs, are delayed from star-formation by stellar evolution (and possibly binary-evolution) timescales, with NS formation coincident with the core-collapse SN, the remaining relevant initial delay is, to zeroeth order, just $t_{i, {\rm gw}}$. Of course,
the stellar and binary evolution timescales of single massive stars may well differ from those of massive binaries that eventually produce DNSs that merge and make kilonovae. However, given our current ignorance about these processes, it is reasonable to simply assume that these timescales are equal for the two kinds of explosion.

As noted before, the maximal value that we consider for $t_i$ (the longest of the three initial timescales that we have defined), $t_i=10$~Myr, is less than the smallest observed value of 
$\tau_{\rm gw}=45$~Myr, and therefore the results of our calculations are largely insensitive to the exact choice of the type of initial delay. In the body of the paper, we hence do not distinguish among them, and denote them all as $t_i$.

\section*{Appendix C\\ Likelihood and MCMC calculation and results}\label{App:MCMC}
The log likelihood of a given  
model distribution, for example a model for the distribution of times until merger,  $dN/d\tau_{\rm gw}$,  in Eqns. \ref{eq.taugwdist1}-\ref{eq.taugwdist2}, is calculated as
\begin{equation}\label{eq:likelihood}
\ln L=\sum_{i=1}^N   \ln \left[\frac{dN}{d\tau_{\rm gw}}(\tau_{{\rm gw}, i}| {\bf\Theta})\right]  , 
\end{equation}
where $\bf \Theta$ is the parameter vector of the relevant model. For a single-power-law DTD model ${\bf\Theta}=\alpha$, with $\alpha$ the power-law index. For a two-power-law DTD model, ${\bf \Theta}=[\alpha_1, \alpha_2, A]$ (the two power-law indices and the relative normalization, see \S\ref{sec:pulsars}). For the cut-off model, ${\bf\Theta}=[\alpha, \tau_{\rm cut}, A]$. The summation is over the $N=18$ DNS systems of the observed sample. This is equivalent, up to a superfluous constant, to calculating the likelihood as the product of the Poisson probabilities, over all values of $\tau_{\rm gw}$, in small bins $\delta \tau_{\rm gw}$, where the observed number of systems in each bin is either one (in the $N$ bins with data) or zero (in all the other bins; see e.g. \citealp{MaozRix1993}). We emphasize that the fit is to the observed distribution of  $\tau_{\rm gw}$ values, rather than to any cumulative or binned distribution (such as those we have shown, for illustrative purposes only, in Figs.~\ref{fig:dN_dtaugw}-\ref{fig:dN_dtaugw2PL}). 

\begin{figure}
    \centering
    \includegraphics[width=8.5cm]{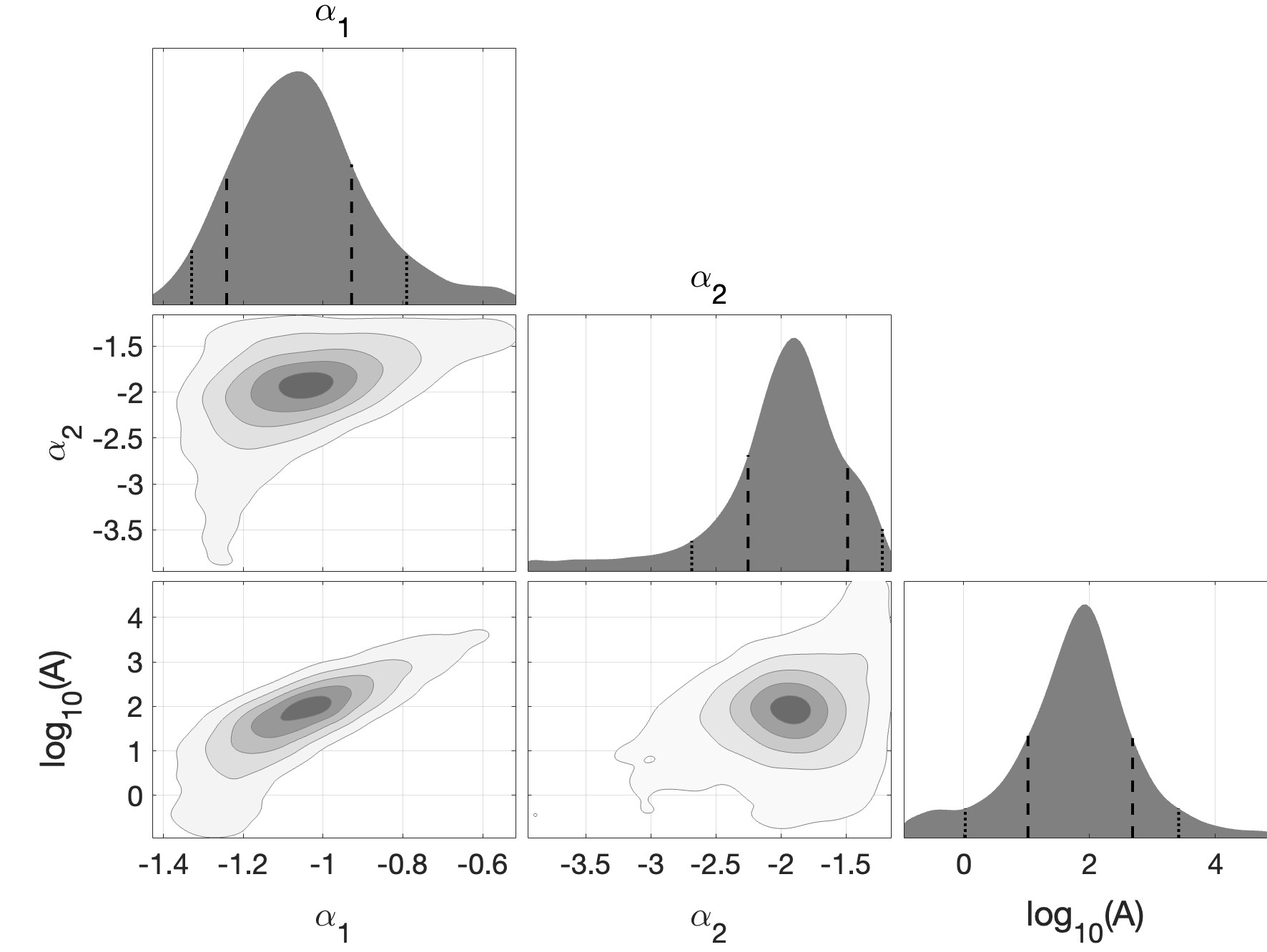}
    \caption{Corner plot of the MCMC maximum likelihood fit (Eq.~\ref{eq:likelihood}) to the observed
    $\tau_{\rm gw}$ values of the sample, using the $dN/d\tau_{\rm gw}$ distribution predicted by a two-power-law DTD model, with indices $\alpha_1$, $\alpha_2$, and relative normalization $A$. Priors on the fit parameters are: $\alpha_1$ distributed uniformly in the range $[-2,-0.5]$; $\alpha_2$ uniformly in the range $[-4,\alpha_1]$; and $\log_{10}(A)$  uniformly in the range $[-1,5]$. The initial DTD delay, $t_i$, is fixed at $10$ Myr. Dashed (dotted) lines marks the regions  containing 68\% (90\%) of the marginalized distribution.} 
    \label{figcorner2PL}
\end{figure}

\begin{figure}
    \centering
    \includegraphics[width=8.5cm]{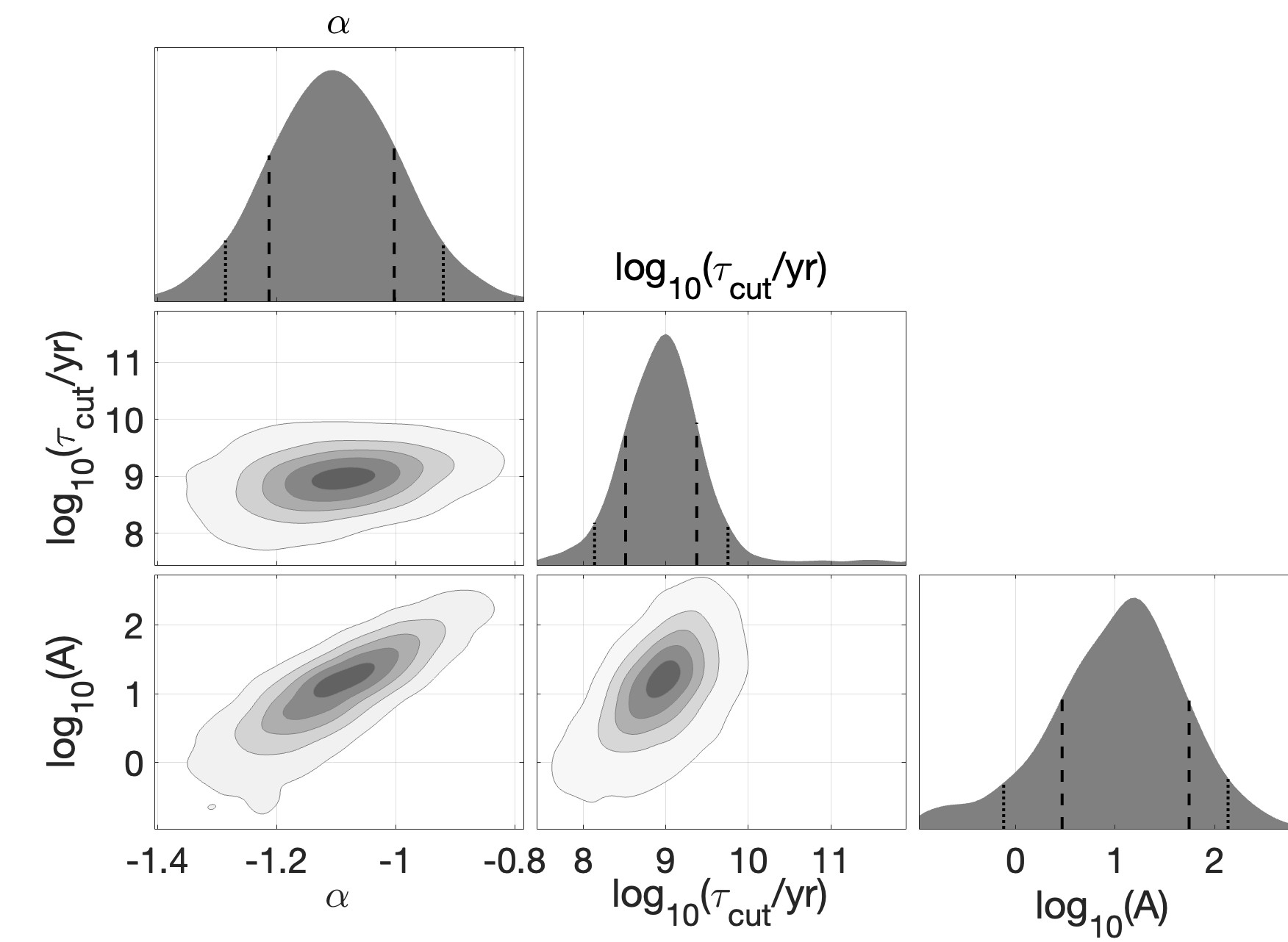}
    \caption{Same as Fig.~\ref{figcorner2PL}, but for the cutoff DTD model, consisting of: one power law with a free-parameter index $\alpha$, another power law with index $-1$ and an exponential cutoff time $\tau_{\rm cut}$, and number ratio $A$. Priors are: $\alpha$ distributed uniformly in the range $[-1.5,-0.5]$, $\log_{10}(\tau_{\rm cut}/{\rm{yr}})$ distributed uniformly in the range $[7,12]$, and $\log_{10}(A)$ distributed uniformly in the range $[-1,5]$. As in Fig.~\ref{figcorner2PL}, $t_i$ is fixed to $10$~Myr.}
    \label{figcorner_cut}
\end{figure}

Using a Markov-Chain Monte-Carlo (MCMC) procedure, we test the likelihood of parametrized families of models,
to find the best-fit parameters for a given family.
Figs.~\ref{figcorner2PL} and \ref{figcorner_cut} show the MCMC "corner plots", showing the model probabilities for combinations of pairs of parameters, for the two-power-law model and for the cut-off models, respectively. For both types of two-population models, we fix the DTD initial delay, $t_i$, at 10 Myr. Varying its value between 1~Myr and 50~Myr affects slightly the best-fit parameters, but the overall results and conclusions remain unchanged. For example, in the two power-law model, varying $t_i$ in the range 1-50 Myr
leaves the best-fit parameter  $\alpha_1$ at $\approx -1.1$, and shifts the best-fit values of the other parameters within the ranges $\alpha_2=-1.75$ to $-2.45$ and $A=60$ to $200$. We do not attempt to constrain $t_i$ based on the observations, as it is possible that
the observed sample at $t_{\rm gw} \lesssim 50$ Myr suffers from observational selection that disfavors detection of rMSPs in DNSs in tight orbits.   

\section*{ORCID iDs}
\noindent Dan Maoz  https:/orcid.org/0000-0002-6579-0483 \\
          Ehud Nakar https:/orcid.org/0000-0002-4534-7089

\bibliographystyle{mnras}
\bibliography{refs_BNS_EU}

\end{document}